\documentclass[largeformat]{interact}

\usepackage{epstopdf}
\usepackage[caption=false]{subfig}

\usepackage[numbers,sort&compress,merge]{natbib}
\bibpunct[, ]{[}{]}{,}{n}{,}{,}

\theoremstyle{plain}

\theoremstyle{definition}

\theoremstyle{remark}

\begin{document}


\title{Rotational spectroscopy of the thioformaldehyde isotopologues H$_2$CS and H$_2$C$^{34}$S 
       in four interacting excited vibrational states and an account on the rotational spectrum of thioketene, H$_2$CCS}

\author{
\name{
Holger S.~P. M\"uller\textsuperscript{a}\thanks{CONTACT Holger S.~P. M\"uller. 
Email: hspm@ph1.uni-koeln.de}\thanks{Supplemental data for this article can be accessed at https://doi.org/10.1080/... .}, 
Atsuko Maeda\textsuperscript{b}, 
Frank Lewen\textsuperscript{a}, 
Stephan Schlemmer\textsuperscript{a}, 
Ivan R. Medvedev\textsuperscript{b,c} and 
Eric Herbst\textsuperscript{b,d}
}
\affil{
\textsuperscript{a}Astrophysik/I. Physikalisches Institut, Universit\"at zu K\"oln, Z\"ulpicher Str. 77, 50937 K\"oln, Germany; 
\textsuperscript{b}Department of Physics, The Ohio State University, Columbus, OH 43210-1107, USA; 
\textsuperscript{c}Department of Physics, Wright State University, Dayton, OH 45435, USA; 
\textsuperscript{d}Departments of Chemistry and Astronomy, University of Virginia, Charlottesville, VA 22904, USA
}
}

\maketitle

\begin{abstract}
An investigation of the rotational spectrum of the interstellar molecule thioformaldehyde between 110 and 377~GHz 
through a pyrolysis reaction revealed a multitude of absorption lines assignable to H$_2$CS and H$_2$C$^{34}$S 
in their lowest four excited vibrational states besides lines of numerous thioformaldehyde isotopologues 
in their ground vibrational states reported earlier as well as lines pertaining to several by-products. 
Additional transitions of H$_2$CS in its lowest four excited vibrational states were recorded in selected regions 
between 571 and 1386~GHz. Slight to strong Coriolis interactions occur between all four vibrational states 
with the exception of the two highest lying states because both are totally symmetric vibrations. 
We present combined analyses of the ground and the four interacting states for our rotational data of H$_2$CS and 
H$_2$C$^{34}$S. The H$_2$CS data were supplemented with two sets of high-resultion IR data in two separate analyses. 
The $v_2 = 1$ state has been included in analyses of Coriolis interactions of low-lying fundamental states of 
H$_2$CS for the first time and this improved the quality of the fits substantially. 
We extended furthermore assignments in $J$ of transition frequencies of thioketene in its ground vibrational state. 
\\\resizebox{22pc}{!}{\includegraphics{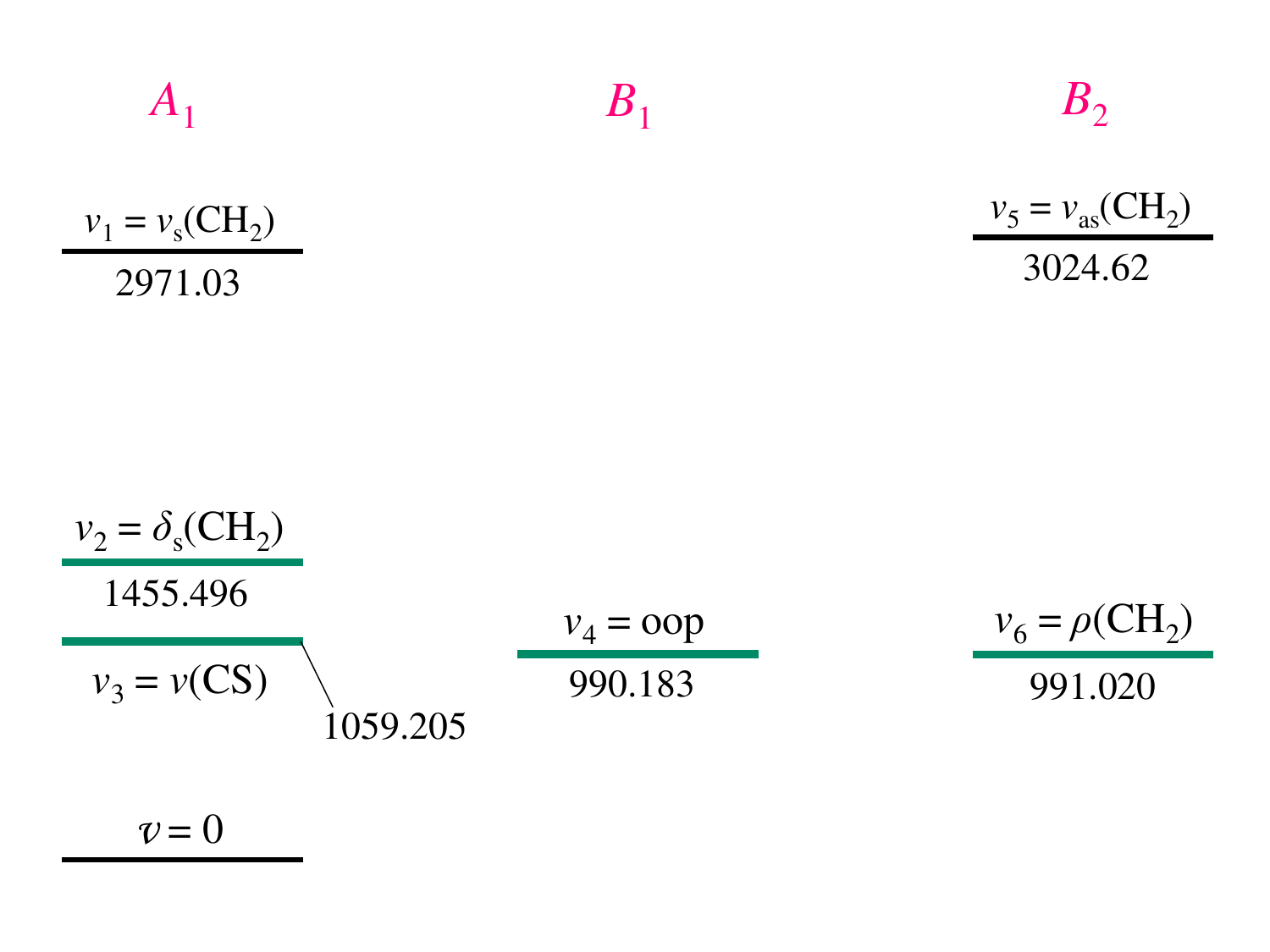}}
\end{abstract}


\section{Introduction}
\label{intro}

Thioformaldehyde, H$_2$CS, is as a small molecule of C\textsubscript{2v} symmetry obviously of fundamental interest 
in particular in comparison to its lighter sibling formaldehyde, H$_2$CO. Its rotational spectrum received additional 
attention because it served as a mean to identify the molecule in a variety of astronomical sources. 
It was detected first in the giant high-mass starforming region Sagittarius B2 near the Galactic centre 
\cite{H2CS_det_1973} and later in dark clouds \cite{H2CS_dark-clouds_1989}, circumstellar envelopes of 
late-type stars \cite{H2CS_etc_CW-Leo_2008}, nearby \cite{LMC_SMC_1999,NGC253_2005} and more distant galaxies 
\cite{PKS1830_4mm_2011} and also in the comet Hale$-$Bopp \cite{H2CS_Hale-Bopp_1997}. 
Several of its isotopologues were also found in space, including H$_2$C$^{34}$S \cite{H2CS-34_1985}, 
H$_2^{13}$CS \cite{SgrB2_survey_1986}, HDCS \cite{HDCS_rot_det_1997} and D$_2$CS \cite{D2CS_det_2005}, 
where unlabelled atoms refer to $^{12}$C and $^{32}$S. Rotational transitions of thioformaldehyde were 
employed more recently to infer temperature in \cite{H2CS_T_in_disks_2020} or the structure of disks around 
young stellar objects \cite{structures_disk-forming_2020} or to investigate deuteration in a prestellar core 
\cite{H2CS-deuteration_L1544_2022}.

The first report on the rotational spectrum of the main isotopologue was published in 1970 \cite{H2CS_rot_1970}. 
Later studies extended the frequency range to 245~GHz or presented data on isotopic species 
\cite{H2CS_isos_rot_1971,H2CS_rot_1972,H2CS_isos_1982,H2CS_HFS_1987}. Transition frequencies of HDCS were determined 
some time later \cite{HDCS_rot_det_1997}, and astronomical observations were employed to improve mainly the 
D$_2$CS data set \cite{D2CS_det_2005}. The dipole moment of H$_2$CS was determined through Stark effect measurements 
\cite{H2CS_isos_rot_1971,H2CS_dip_1977}. All these studies were restricted to the ground vibrational states.

Medium- and high-resolution IR investigations were performed in the CH$_2$ stretching region covering $\nu _5$, $\nu _1$ 
and $2\nu _2$ \cite{H2CS_nu1_nu5_2nu2_1971}, in the $\nu _2$ CH$_2$ bending region \cite{H2CS_FIR_nu2_1993} and around 
10~$\umu$m \cite{H2CS_laser-Stark_1980,H2CS_D2CS_IR_re-est_1981,H2CS_IR_2008}. The 10~$\umu$m studies include 
laser Stark spectroscopy of $\nu _4$, $\nu _6$ and $\nu _3$ with determination of permanent dipole moments 
in the excited vibrational states \cite{H2CS_laser-Stark_1980}, an FTIR investigation of $\nu _4$ and $\nu _6$ 
of H$_2$CS with additional results on D$_2$CS, a harmonic force field calculation and an estimate of the equilibrium 
structure of thioformaldehyde \cite{H2CS_D2CS_IR_re-est_1981} and finally an extensive FTIR study of $\nu _4$, 
$\nu _6$ and $\nu _3$ \cite{H2CS_IR_2008}.

Thioformaldehyde has also been subjected to several studies of its electronic spectrum. Most of these were, however, 
of limited impact for the data in the ground electronic state. A notable exception is a sub-Doppler spectroscopic 
investigation of a part of the $\tilde{A} - \tilde{X}$ spectrum that yielded ground state combination differences which 
improved the purely $K$-depended parameters considerably \cite{H2CS_A-X_1994}.

There have also been numerous quantum-chemical calculations on structural or vibrational properties of H$_2$CS, 
in particular in the past 30 years 
\cite{H2CS_FF_ai_1994,H2CS_ai_1998,H2CS_vib_ai_2006,H2CS_ai_2011,H2CS_ai_egy-etc_2013,H2CS_rot_2019}. 
The existing experimental rotational and rovibrational data were evaluated recently \cite{H2CS_marvel_2023}, 
and the results were employed in a refinement of a quantum-chemically generated potential energy surface 
to generate a high-temperature line list of H$_2$CS \cite{H2CS_ExoMol_2023}. We mention furthermore calculations 
of the properties of isomers of thioformaldehyde \cite{H2CS_isomerization_1999,H2CS_isomerization_2020}. 
Finally, H$_2$CS served also as an example to improve \cite{improve_mm_2009} or develop programs to 
calculate line intensities \cite{calc-int_H2CS_2020} or spectroscopic parameters \cite{calc-BDH_2022}.

Large uncertainties in the transition frequencies of the H$_2$CS main isotopologue in the submillimetre region 
prompted us to investigate the rotational spectrum of thioformaldehyde at The Ohio State University and 
subsequently at the Universit\"at zu K\"oln resulting in a report on the main isotopic species \cite{H2CS_rot_2008}. 
Measurements at the Universit\"at zu K\"oln were extended in frequency some time later leading to a 
further improved account on the spectra of the main isotopologue and data pertaining to several minor 
isotopic species observed in natural isotopic composition, including the very rare H$_2^{13}$C$^{34}$S, 
HDCS and H$_2$C$^{36}$S \cite{H2CS_rot_2019}. The wealth of accurate rotational parameters of many isotopic species 
was taken to evaluate a semi-empirical equilibrium structure with vibration$-$rotation parameters from 
quantum-chemical calculations \cite{H2CS_rot_2019}. The measurements carried out at The Ohio State University 
covered large sections of the millimetre wave and the lower part of the submillimetre wave region continuously 
and revealed many more absorption features besides those of several thioformaldehyde isotopologues in their 
ground vibrational states. A large amount of these lines could be assigned to transitions of H$_2$CS and H$_2$C$^{34}$S 
in their lowest four excited vibrational states. Other lines could be assigned to by-products of the pyrolysis reaction 
through which thioformaldehyde was generated. Our present and final work on the rotational spectroscopy of thioformaldehyde 
deals with analyses of the Coriolis-coupled lowest four excited vibrational states of H$_2$CS and H$_2$C$^{34}$S. 
These states are in fact textbook examples of Coriolis coupling because $v_4 = 1$ and $v_6 = 1$ are essentially degenerate, 
$v_3 = 1$ is quite close to these two states while $v_2 = 1$ is more distant, but still close enough that it is necessary 
to consider this state in the analyses. 
We supplement our H$_2$CS analyses by two sets of high-resolution data of $\nu _4$, $\nu _6$ and $\nu _3$ and one set 
of data of $\nu _2$. We present also a reanalysis of the ground state rotational spectrum of thioketene, H$_2$C$_2$S, 
one of the by-products of the pyrolysis reaction, whose rotational spectrum was presented up to 226~GHz in 
the previous literature \cite{H2CCS_rot_1979a,H2CCS_rot_1979b,H2CCS_rot_1980}. 
It is worthwhile mentioning that its ground state spectroscopic parameters were also improved in a far- and mid-IR 
spectroscopic study \cite{H2CCS_IR_1996} and that it was detected in the cold and dense prestellar core TMC-1 
quite recently besides several other sulfur-containing molecules \cite{det_H2C2S_etc_2021}.

The rest of this manuscript is organised as follows. Section~\ref{exptl} provides details on our laboratory measurements. 
The spectroscopic properties of thioformaldehyde are described in Section~\ref{spec-prop} while Section~\ref{sec_fitting} 
deals with considerations for the analyses and the fitting of the spectra. Our results are detailed in Section~\ref{results}, 
discussed in Section~\ref{discussion} and concluding remarks are presented in Section~\ref{conclusion}.

\section{Laboratory spectroscopic details}
\label{exptl}

The Fast Scan Submillimetre-wave Spectroscopic Technique (FASSST) was developed at The Ohio State University (OSU) 
and employed there to cover most of the 110$-$377~GHz range \cite{FASSST_1997,DEE_rot_FASSST_2004,sub-mmW_2010}. 
We used furthermore two different spectrometer systems at the Universit{\"a}t zu K{\"o}ln to record higher 
frequency transitions up to almost 1.4~THz \cite{THz-BWO_1994,THz-BWO_1995,MeSH_rot_2012}.

The FASSST system applies backward wave oscillators (BWOs) as sources; in the present investigation one 
that covers about 110$-$190~GHz and two additional ones spanning the region of 200$-$377~GHz. 
The frequency of each BWO is swept very quickly such that a wide frequency range ($\sim$90~GHz) 
can be measured in a short period and any voltage instability of the BWOs can be overcome. 
Each FASSST spectrum requires calibration which was achieved through rotational lines of SO$_2$. 
It displays sufficiently many spectral frequencies which are well known \cite{SO2_rot_2005}. 
A portion of the source radiation propagates through a Fabry$-$Perot cavity to produce an interference fringe 
spectrum with a free spectral range of $\sim$9.2~MHz. The frequencies of radiation between the calibration 
lines are interpolated with the fringe spectrum. 
It is important to take the dispersive effect of atmospheric water vapour in the Fabry$-$Perot cavity 
into account in the calibration procedure \citep{dispersion-corr_2006,dispersion-corr_2007}. 
Measurements were taken with scans that proceeded both upward and downward in frequency to record an average frequency. 
The results obtained from 100 upward and downward scans were accumulated for a better signal-to-noise ratio (S/N), 
increasing the integration time from $\sim$0.1 to $\sim$10~ms per Doppler limited line width. 
The experimental uncertainty of this apparatus is around 50~kHz for an isolated, well-calibrated line.

The Cologne spectrometers are equipped with phase-lock loop (PLL) systems to obtain accurate frequencies. 
Two BWOs were employed as sources to record usually individual lines in the 566$-$670 and 848$-$930~GHz regions. 
A portion of the radiation from the BWOs is mixed with an appropriate harmonic of a continuously tunable synthesiser 
in a Schottky diode multiplier mixer to produce the intermediate frequency (IF) signal. The IF signal is phase locked 
and the phase error provided by the PLL circuit is fed back to the power supply of the BWOs. Further details on this 
spectrometer system are available elsewhere \cite{THz-BWO_1994,THz-BWO_1995}.

Virginia Diode Inc. (VDI) frequency multipliers driven by an Agilent E8257D microwave synthesiser were used 
to record transition frequencies between 1290 and 1390~GHz \cite{MeSH_rot_2012}. Both spectrometer systems 
achieve accuracies of 10~kHz and even better for very symmetric lines with good S/N as shown in recent 
studies on vibrationally excited methyl cyanide \cite{MeCN_up2v4eq1_etc_2021} or on isotopic oxirane 
\cite{c-C2H4O_rot_2022}.

Thioformaldehyde (H$_2$CS) was generated by the pyrolysis of trimethylene sulfide [(CH$_2$)$_3$S; Sigma-Aldrich Co.], 
which was used as provided. The thermal decomposition of trimethylene sulfide affords thioformaldehyde and ethylene 
in high yields. Ethylene does not have a permanent dipole moment so its presence is essentially negligible. 
Small amounts of other by-products are present in the spectrum and include CS, H$_2$S and H$_2$CCS. 
Laboratory setups for the pyrolysis were slightly different in the OSU and Cologne measurements. 
At OSU, trimethylene sulfide vapour was passed through a 2-cm diameter, 20~cm long piece of quartz tubing stuffed with 
quartz pieces and quartz cotton to enlarge the reaction surface. The quartz tubing was heated with a cylindrical furnace 
to $\sim$680$^{\circ}$C. The gas produced from the pyrolysis was introduced to a 6-m-long aluminum cell at room temperature 
and pumped to a pressure of 0.4$-$1.5~mTorr (1~mTorr = 0.1333~Pa). The spectrum of trimethylene sulfide disappeared 
almost totally after the pyrolysis, at which time the spectrum of thioformaldehyde appeared. Spectral lines of by-products 
were usually less intense compared with those of thioformaldehyde.

A 3-m-long glass absorption cell kept at room temperature was used for measurements at Cologne. 
A higher temperature of about 1300$^{\circ}$C was required in the pyrolysis zone in order to maximise 
the thioformaldehyde yield and to minimise absorptions of (CH$_2$)$_3$S because no quartz cotton was 
used in the quartz pyrolysis tube. The total pressure was around 1$-$3~Pa for weaker lines and 
around 0.01$-$0.1~Pa for stronger lines.

Liquid He-cooled InSb bolometers were used in both laboratories as detectors. Frequency modulation was 
employed at Cologne to reduce baseline effects. The demodulation at twice the modulation frequency causes 
absorption lines to appear approximately as second derivatives of a Gaussian. 
Spectral baselines in OSU spectra were reduced by filtering of detector signals produced by fast scan 
of the radiation source through the spectral line. The detected effective line shape is near first derivative. 
Numerical differentiation leads to near second derivative line shapes. Additional digital filtering 
suppresses the baseline further.

\section{Spectroscopic properties of thioformaldehyde}
\label{spec-prop}

Thioformaldehyde is an asymmetric rotor with $\kappa = (2B - A - C)/(A - C) = -0.9924$ very close to 
the symmetric limit of $-1$. Its dipole moment of 1.6491~D \citep{H2CS_dip_1977} is aligned with the $a$ 
inertial axis. The strong rotational transitions are therefore those with $\Delta K_a = 0$ and $\Delta J = +1$ 
called $R$-branch transitions. Transitions with $\Delta K_a = 0$ and $\Delta J = 0$ ($Q$-branch transitions) 
are also allowed as are transitions with $\Delta K_a = \pm2$. But these transitions are much weaker 
than the strong $R$-branch transitions because of the proximity of $\kappa$ to $-1$ and 
none have been identified for excited vibrational states.

Isotopologues with two H (and also those with two D) have C\textsubscript{2v} symmetry. 
Three of the six fundamental vibrations are in the totally symmetric symmetry class $A_1$, one is in 
$B_1$ and two are in $B_2$ with transition dipole moments $\mu _a$, $\mu _c$ and $\mu _b$ respectively. 
The two equivalent H nuclei in H$_2$CS and H$_2$C$^{34}$S lead to \textit{ortho} and \textit{para} 
spin-statistics with a 3~:~1 weight ratio. The \textit{ortho} states are described by $K_a$ being odd 
in vibrational states of $A$ symmetry whereas $K_a$ is even in $B$ symmetry states.

The origins of the CH$_2$ stretching states $v_1 = 1$ and $v_5 = 1$ are at 2971.03 and 3024.62~cm$^{-1}$, 
respectively, for the H$_2$CS main isotopologue \cite{H2CS_nu1_nu5_2nu2_1971}. Rotational transitions 
within these states are weaker than a factor of 10$^{-6}$ at 300~K compared with the ground vibrational states 
and hence unobservable in normal absorption spectra. Rotational transitions of the remaining fundamental 
vibrational states have been identified in the course of our investigations. The state $v_2 = 1$ at 
1455.496~cm$^{-1}$ is the CH$_2$ bending state while $v_3 = 1$ at 1059.205~cm$^{-1}$ is the CS stretching state. 
Their Boltzmann factors at 300~K are $9.3 \times 10^{-4}$ and $6.2 \times 10^{-3}$, respectively, while those 
of the out-of-plane state $v_4 = 1$ at 990.183~cm$^{-1}$ and of the CH$_2$ rocking state $v_6 = 1$ at 991.020~cm$^{-1}$ 
are about $8.6 \times 10^{-3}$. The near-degeneracy of two states with one more state close by and another one 
somewhat more distant make the low-lying fundamentals of thioformaldehyde a textbook example of Coriolis coupling. 
The interactions of $v_4 = 1$ with $v_2 = 1$ or $v_3 = 1$ obey $b$-type selection rules, meaning that $\Delta J = 0$, 
$\Delta K_a$ and $\Delta K_c$ are odd. The interactions of $v_6 = 1$ with $v_2 = 1$ or $v_3 = 1$ follow 
$c$-type selection rules with $\Delta J = 0$, $\Delta K_a$ odd and $\Delta K_c$ even. No interactions occur 
between $v_2 = 1$ and $v_3 = 1$ because they are in the same symmetry class while $a$-type selection rules 
govern the interactions between $v_4 = 1$ and $v_6 = 1$.

\begin{figure}
\centering
\resizebox*{15cm}{!}{\includegraphics{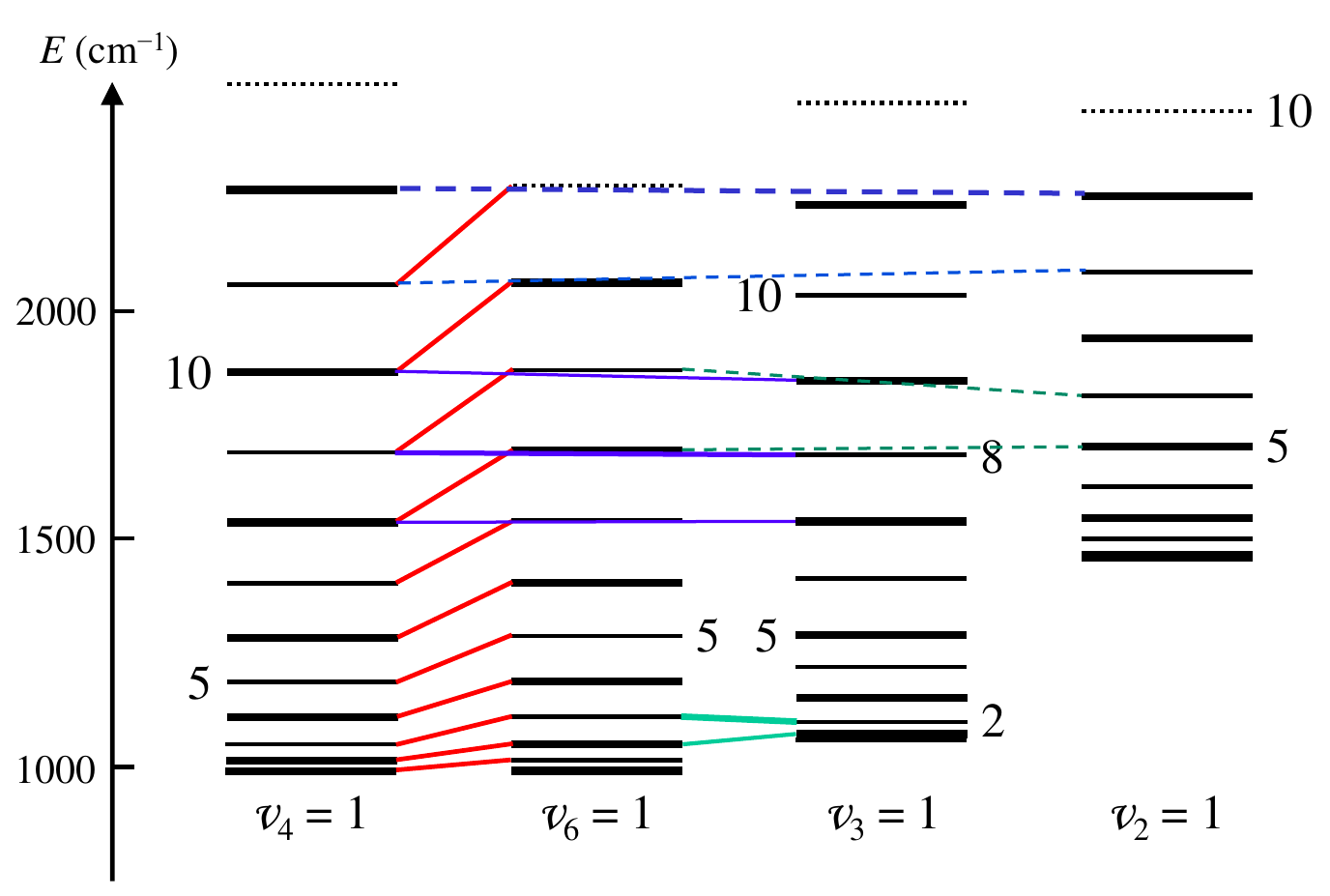}}
\caption{Diagram showing the $K_a$ energy level of the four lowest excited vibrational states of H$_2$CS. 
   Black solid lines indicate levels accessed in the present investigation, dotted ones the first level 
   not accessed; \textit{ortho} levels are indicated by thicker lines. Please not that the lines for 
   $K_a = 0$ and 1 are often not separated. Red, blue and green lines signal (near-) resonant 
   $a$-, $b$- and $c$-type Coriolis interactions, respectively. Solid lines stand for $\Delta K_a = 0$ 
   or 1 interactions, dashed lines for $\Delta K_a = 3$ interactions.} 
\label{K-energy-level}
\end{figure}

Figure~\ref{K-energy-level} displays the $K_a$ level structure of $v_4 = 1$, $v_6 = 1$, $v_3 = 1$ 
and $v_2 = 1$ from left to right with \textit{ortho}-levels marked by thicker lines; (near-) resonant 
Coriolis interactions are indicated with coloured lines. Coriolis interactions in a strict sense 
are interactions between fundamental vibrational states as well as interactions between vibrational states 
where a particular number of vibrational quanta have been added to both of these interacting vibrational states. 
Equivalent interactions in a more general sense are called rotational resonances because the strength of 
the interaction scales with $J$ and $K$.

The most striking feature in the $K_a$ energy level diagram is the $a$-type interaction between the two 
essentially degenerate states $v_4 = 1$ and $v_6 = 1$. The $K_a$ levels repel each other with increasing 
$K_a$ starting from $K_a = 1$ as the selection rules involve $\Delta K_a = 0$. 
The asymmetry splitting causes the interaction between the upper asymmetry level of a particular $K_a$ 
of $v_4 = 1$ with the lower asymmetry level of the same $K_a$ of $v_6 = 1$ to be stronger than the 
interaction between the lower asymmetry level of $v_4 = 1$ with the upper asymmetry level of $v_6 = 1$ 
because the latter pair of levels is farther apart than the former. 
The asymmetry splitting in $v_4 = 1$ is inverted in frequency and in energy at lower values of $J$ as 
a consequence and both $K_a = 1$ levels of $v_4 = 1$ are lower in energy than $K_a = 0$ up to $J = 6$. 
The assignments to the upper asymmetry level of $v_4 = 1$ and the lower asymmetry level of $v_6 = 1$ 
swap at some higher value of $J$. The asymmetry splitting in an unperturbed vibrational state 
is largest in $K_a = 1$ so this swap in assignment occurs in the H$_2$CS isotopologue already at $J = 7$, 
at $J = 23$ for $K_a = 2$ and at $J = 44$ for $K_a = 3$.

The $v_3 = 1$ state is perturbed through $c$-type Coriolis interaction with $v_6 = 1$ at low values of $K_a$ 
with $K_a = 2$ of $v_3 = 1$ being very close to $K_a = 3$ of $v_6 = 1$ in particular at low values of $J$, see 
Fig.~\ref{K-energy-level}; $K_a = 1$ of $v_3 = 1$ get very close to $K_a = 2$ of $v_6 = 1$ at higher values 
of $J$ with a resonant $\Delta K_c = 2$ crossing in energy between $J = 39$ and 40. More resonant crossings 
occur beyond the range of $J$ values covered in the present study. 
Perturbations of higher values in $K_a$ of $v_3 = 1$ occur through $\Delta K_a = 1$ $b$-type Coriolis interaction 
with $v_4 = 1$ and $K_a = 7$ to 9 of $v_3 = 1$ interacting most strongly with $K_a = 8$ to 10 of $v_4 = 1$. 
A resonant crossing in energy occurs for $K_a = 8$ of $v_3 = 1$ and $K_a = 9$ of $v_4 = 1$ between $J = 30$ 
and 31.

The vibrational energy of $v_2 = 1$ is substantially higher than those of the lowest three excited vibrational states. 
Interactions with $v_4 = 1$ and $v_6 = 1$ involve near-degeneracies of $b$- and $c$-type selection rules with 
$\Delta K_a = 3$ as is demonstrated in Fig.~\ref{K-energy-level}. Levels with $K_a = 9$ of $v_2 = 1$ and $K_a = 12$ of 
$v_4 = 1$ are somewhat close in energy while $K_a = 8$ of $v_2 = 1$ and $K_a = 11$ of $v_4 = 1$ are less close in energy. 
Levels with $K_a = 5$ of $v_2 = 1$ and $K_a = 8$ of $v_6 = 1$ as well as those with $K_a = 6$ of $v_2 = 1$ and $K_a = 9$ 
of $v_6 = 1$ are also comparatively close in energy.

\section{Analysis and fitting}
\label{sec_fitting}

Assignments in the OSU FASSST spectra of thioformaldehyde were carried out with the Computer Aided 
Assignment of Asymmetric Rotor Spectra (CAAARS) program applying the Loomis$-$Wood procedure, 
with which an observed spectrum is visually compared with calculated line positions and intensities 
to make new assignments \cite{CAAARS_2005}. The software facilitates user-defined sorting of predicted 
transitions into branches. It then overlays consecutive transitions into a Loomis$-$Wood diagram, 
thus, aiding visual search for matching lines between predictions and experiment.

Calculation and fitting of spectra were carried out with Pickett's SPCAT and SPFIT programs \cite{spfit_1991}. 
It is important to apply the assignment option ''eigenvector sort of states'' instead of the default 
''energy sort of Wang sub-blocks'' in cases with pronounced first-order Coriolis coupling. 
An informative example is the $v_8 = 3$ $l = +3$ substate in which $K$ levels decrease from $K = 0$ 
because of the strong Coriolis interaction with the $l = -3$ substate to $K = 3$ and only $K \geq 6$ 
are higher in energy than $K = 0$ \cite{MeCN_nu4_nu7_3nu8_1993}. 
The option ''eigenvector sort of states'' secures this labelling. Additional aspects of the labelling 
of $K$ quantum numbers in SPCAT and SPFIT are available elsewhere \cite{spfit_1991}.

Watson's S reduction was employed in the rotational Hamiltonian. It is more versatile in general 
and obviously more appropriate than the A reduction in the case of thioformaldehyde because of 
the proximity to the symmetric prolate limit. The centrifugal distortion parameters in an excited 
vibrational state of a fairly rigid molecule are often quite close to those in the ground vibrational 
state. We apply the ground state rotational parameters $X_0$ of H$_2$CS or of H$_2$C$^{34}$S to all 
excited vibrational states and introduce differences $\Delta X_i = X_i - X_0$ as far as needed; 
$i$ represents an excited vibrational state. This approach has several advantages of which the most 
important ones are first the ground state centrifugal distortion parameters account commonly to a considerable 
amount for the distortion effects in an excited vibrational state and second it is easier to recognise 
if a particular $\Delta X_i$ is determined with sufficient significance. Both aspects may help to reduce 
the number of spectroscopic parameters to reach a satisfactory fit. And finally large parameters 
$\Delta X_i$ and $\Delta X_j$ of two vibrational states of similar magnitudes and opposite signs 
may indicate an unaccounted perturbation.

To keep the parameter set small and somewhat unique we test after each round of assignments if 
the use of a particular parameter improves the rms error of the fit as a measure of the quality of the fit. 
We search among the meaningful parameters for the one that improves the rms error most. The parameter is kept 
in the fit if the improvement is deemed to be sufficient and if the parameter is determined with significance. 
The procedure is repeated until improvements of the fit are marginal at most. This fitting strategy works 
usually very well for molecules close to the prolate symmetric limit, but occasionally less well for very 
asymmetric rotors or rotors closer to the oblate symmetric limit.

The vibration$-$rotation interaction between two vibrational states is commonly treated with a Hamiltonian 
that can be divided into a $2 \times 2$ matrix with the diagonal elements consisting usually of two Watson-type 
rotational Hamiltonians, including the vibrational energy of each state, and the interaction Hamiltonian 
off-diagonal. This procedure is extended analogously in cases of several interacting vibrational states.

The low order Coriolis terms of $a$-symmetry are 
\[ 
iG_a J_a + F_{bc}\{J_b,J_c\}
\] 
with $\{,\}$ being the anticommutator; terms of $b$- and $c$-symmetry are defined equivalently. 
We point out that other designations than $G$ and $F$ may be found, in particular in the older literature. 
One of the early and well-recognised examples to establish that both terms are not only allowed but also needed 
was a study of the $c$-symmetry Coriolis interaction between $v_1 = 1$ and $v_3 = 1$ in ozone \cite{Coriolis_1970}. 
These terms may be supplemented with distortion corrections of the type
\[ 
i(G_{a,K}J_a^3 + G_{a,J}\{J_a,J^2\} + G_{2a}\{J_a,J_b^2 - J_c^2\} + ...) 
\] 
and 
\[ 
F_{bc,K}\{J_a^2,\{J_b,J_c\}\} + F_{bc,J}J^2\{J_b,J_c\} + F_{2bc}(J_b^2 - J_c^2)\{J_b,J_c\} + ... 
\] 
in the case of a prolate representation. Such terms were required extensively for example in the treatment 
of the Coriolis interaction of $v_4 = 1$ and $v_6 = 1$ in ClClO$_2$ \cite{ClClO2_Coriolis_2002}. 
The sign of $G_a$ coupling two specific vibrations is usually not determinable through a fit. Its sign may, 
however, affect the intensities of some rotational or rovibrational transitions in conjunction with the signs 
of permanent or transition dipole moment components. The signs of $F_{bc}$ or of distortion corrections to 
$G_a$ or $F_{bc}$ are determinable in fits relative to the sign of $G_a$.

In the case of Coriolis interaction between two fundamental vibrations 
$\nu _x$ and $\nu _y$ the associated $G_i(x,y)$ can be evaluated through 
\[
G_i(x,y) = \zeta ^i_{x,y} B^i_e \left(\sqrt{\frac{\omega _x}{\omega _y}} + \sqrt{\frac{\omega _y}{\omega _x}}\right)
\]
where $\omega _x$ and $\omega _y$ are the corresponding harmonic vibrations, $B^i_e$ is the 
$i$-axis equilibrium rotational parameter and $\zeta ^i_{x,y}$ a Coriolis term that can be 
evaluated from a harmonic force field calculation. 
Replacing $\omega _x$ and $\omega _y$ with $\nu _x$ and $\nu _y$ has usually only a small effect 
whereas the substitution of $B^i_e$ by $B^i_0$ may lead to a more pronounced change. 
Vibration$-$rotation interaction frequently increases correlation in the fitting procedure. 
Fixing one or more $G_i$ to values from a force field calculation is a way to reduce correlation 
in cases of Coriolis interaction and results usually in physically more meaningful values for the 
remaining parameters compared to cases in which the $G_i$ are floated or kept fixed at zero. 
Fixing to values derived from a force field calculation is also advisable if the experimental data set 
does not reach the quantum number range of strongest interactions, usually for levels of the 
interacting vibrations with $\Delta K_a = 0$ for $a$-type interaction and $\Delta K_a = 1$ for 
$b$- or $c$-type interaction in case of a prolate rotor. This was done for example for $G_a$ and $G_b$ 
in the case of $v_4 = 1$ and $v_6 = 1$ in ClClO$_2$ \cite{ClClO2_Coriolis_2002} or for $G_c$ in 
the case of $v_1 = 1$ and $v_3 = 1$ in SO$_2$ \cite{SO2_Triade1_2013}.

We carried out quantum-chemical calculations at the Regionales Rechenzentrum der Universit{\"a}t 
zu K{\"o}ln (RRZK) to evaluate $\zeta ^i_{x,y}$ and $\nu _x$  for H$_2$CS and H$_2$C$^{34}$S. 
We applied MP2 M{\o}ller$-$Plesset perturbation theory of second order \cite{MPn_1934} using a 
correlation-consistent basis set of quadruple zeta quality augmented with diffuse and tight 
core-correlating basis functions designated as aug-cc-pwCVQZ \cite{cc-pVXZ_1989,core-corr_2002}. 
We resorted to the commercially available program Gaussian~16 \cite{Gaussian16C} and applied 
the default frozen core option.

\section{Results}
\label{results}

We describe in the following our results obtained for the four lowest excited vibrational states 
of H$_2$CS and H$_2$C$^{34}$S and the contribution to the ground vibrational state of H$_2$C$_2$S.

\subsection{H$_2$CS}
\label{H2CS-results}

The Loomis$-$Wood display of the OSU spectral recordings revealed several unassigned series in $K_a$ 
that were slightly weaker than those of H$_2^{13}$CS and that resembled more or less those of H$_2$CS 
after having assigned transitions of several thioformaldehyde isotopologues \cite{H2CS_rot_2019}. 
It was quite natural to assume these are caused by low-lying excited vibrational states of H$_2$CS. 
This assumption was easily verified applying spectroscopic parameters from two older IR studies in which 
combined analyses of $\nu _4$, $\nu _6$ and $\nu _3$ were presented and the Coriolis interaction 
was analysed as far as it is present in their data \cite{H2CS_laser-Stark_1980,H2CS_D2CS_IR_re-est_1981}.  
Transitions within $v_2 = 1$ could be assigned subsequently based on an IR investigation of $\nu _2$ 
that treated the band as unperturbed despite indications to the contrary \cite{H2CS_FIR_nu2_1993}.

The OSU measurements cover $J = 4 - 3$ to $10 - 9$ or $11 - 10$ encompassing the $J = 7 - 6$ transitions 
between $v_4 = 1$ and $v_6 = 1$ having $K_a = 1$ and $K_c = J - K_a$ in the case of $v_4 = 1$ (the upper 
asymmetry component in an unperturbed vibrational state) and $K_c = J - K_a +1$ in the case of $v_6 = 1$ 
(the lower asymmetry component). The associated transitions within $v_4 = 1$ or $v_6 = 1$ were not calculated 
at a threshold more than a factor of 1000 lower. The assignments extended to $K_a = 8$ for $v_4 = 1$ and 
$v_6 = 1$, to $K_a = 7$ for $v_3 = 1$ and to $K_a = 6$ for $v_2 = 1$. 
Uncertainties of 50~kHz were attributed to all of the excited state lines of H$_2$CS from OSU.

It may be useful to describe qualitatively how the rotational of H$_2$CS in low-lying excited vibrational 
states appear in comparison to the ground vibrational state. We describe the general features of the 
$J = 10 - 9$ transition as it was contained in the OSU recordings. The $K_a = 3$ to 9 lines of $v = 0$ 
cover sequentially from $\sim$343.4~GHz down to 342.6~GHz with the $K_a = 2$ lines to either side of 
the $K_a = 3$ lines and the $K_a = 1$ lines near 338.1 and 348.5~GHz. The appearance of $v_2 = 1$ 
is quite similar. The transitions of $v_4 = 1$ and $v_6 = 1$ occur in much smaller frequency regions. 
The $K_a = 2$ to 8 lines of $v_4 = 1$ occur sequentially from about 343.3 to 341.5~GHz with $K_a = 0$ 
and 1 near $K_a = 6$ and $K_a = 9$ more than 3~GHz lower than $K_a = 8$. The $K_a = 3$ to 9 lines of 
$v_6 = 1$ cover sequentially essentially the same region as most of the $v_4 = 1$ lines, one of the 
$K_a = 1$ lines slightly higher and the remaining low-$K_a$ lines distributed among the higher $K_a$ ones. 
Finally, the $v_3 = 1$ lines occur in a very irregular way. The $K_a = 1$ lines are found at the upper 
and lower end, as one would expect, separated by almost exactly 10~GHz; $K_a = 9$ appears roughly 2~GHz 
above the lower of the $K_a = 1$ lines and $K_a = 8$ roughly 2~GHz below the upper $K_a = 1$ line.

\begin{figure}
\centering
\resizebox*{7cm}{!}{\includegraphics{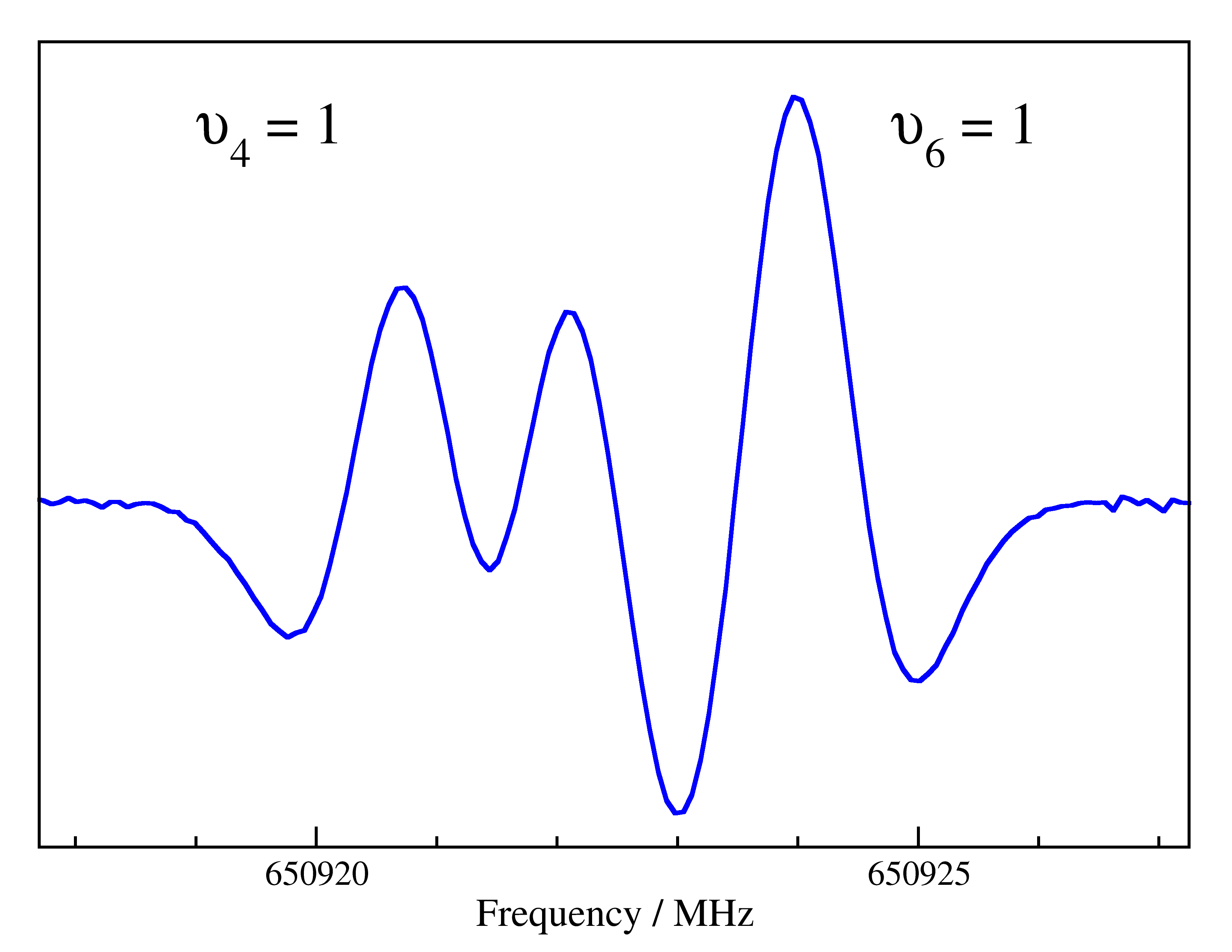}}
\caption{Spectral recording of H$_2$CS displaying the $J = 19 - 18$ transitions with $K_a = 4$ of 
   $v_4 = 1$ and $v_6 = 1$. The $v_4 = 1$ splitting is inverted whereas the $v_6 = 1$ splitting is not 
   resolved.} 
\label{K4-asymmetry}
\end{figure}

After having fitted the OSU excited vibrational state data of H$_2$CS well it was straightforward 
to locate transitions at higher frequencies in Cologne. The BWO measurements between 566 and 670~GHz 
and between 848 and 930~GHz cover $J = 17 - 16$ to $19 - 18$ and $J = 25 - 24$ to $27 - 26$ for as many 
$K_a$ values of the four excited vibrational states as possible. The extent of the coverage differs for 
various reasons. One very obvious reason is that the intensity drops with increasing $K_a$, though 
moderated by the spin statistics, and with vibrational energy. Further reasons are blending with other lines, 
the proximity of much stronger lines, the drop in sensitivity of the BWO or difficulty to lock the BWO. 
The $K_a$ quantum numbers reach 12 (11) for $v_4 = 1$, 10 (8) for $v_6 = 1$, 11 for $v_3 = 1$ and 
9 (7) for $v_2 = 1$ where the numbers in parentheses refer to the 848$-$930~GHz region, see also 
Fig.~\ref{K-energy-level}. The different asymmetry splitting in $K_a = 4$ of $v_4 = 1$ and $v_6 = 1$ 
is shown in Fig.~\ref{K4-asymmetry} for $J = 19 - 18$. The $v_4 = 1$ asymmetry splitting is inverted 
meaning that the transition with $K_c = J + K_a$ is below the one with $K_c = J + K_a + 1$ whereas 
it is opposite in unperturbed vibrational states. The asymmetry splitting in $v_4 = 1$ is $-$1.29~MHz 
(the minus sign indicates the inverted splitting) whereas the unresolved splitting in $v_6 = 1$ is 
calculated to be 0.09~MHz. The observed or calculated values in $v = 0$, $v_3 = 1$ and $v_2 = 1$ 
are 1.31, 1.13 and 1.79~MHz, respectively, for comparison purpose.

Measurements in the 1290$-$1390~GHz region were carried out some time later. They cover $J = 37 - 36$ to 
$41 - 40$. No attempts were made to record transitions of $v_2 = 1$ in the upper frequency region because 
the Boltzmann peak of the rotational spectrum of thioformaldehyde is near 820~GHz at 300~K and because 
of the lower power of the source compared to the BWOs. Nevertheless, $K_a = 8$ was reached for $v_4 = 1$ 
with no coverage of the \textit{para} levels with $K_a = 5$ and 7. Levels with $K_a \leq 3$ and $K_a = 6$ 
were covered in $v_6 = 1$ and levels with $K_a \leq 3$ and $K_a = 7$ were covered in $v_3 = 1$. 
We obtained in particular transition frequencies for $J = 40 - 39$ transitions between $v_6 = 1$ and 
$v_3 = 1$ having $K_a = 2$ and $K_c = J - K_a$ in $v_6 = 1$ and $K_a = 1$ and $K_c = J - K_a + 1$ in 
$v_3 = 1$. The associated transitions within $v_6 = 1$ or $v_3 = 1$ were calculated to be around 
a factor of 50 weaker, too weak to be observed in a reasonable time and possibly difficult to identify 
because of the huge amount of lines of similar or larger intensities. 
Other resonant crossings in energy mentioned in Section~\ref{spec-prop} were in measurement gaps of 
the Cologne measurements or beyond the upper frequency limit. 
Uncertainties of 5 to 10~kHz were assigned to very symmetric lines with good or very good S/N, less 
symmetric lines or lines with somewhat lower S/N were deemed to be more uncertain up to 70~kHz.

The excited state data of H$_2$CS from OSU and from the Universit\"at zu K\"oln were fitted together with 
the previously published ground state data \cite{H2CS_rot_2019}, as well as transition frequencies for 
$\nu _4$ and $\nu _6$ \cite{H2CS_laser-Stark_1980,H2CS_D2CS_IR_re-est_1981}, $\nu _3$ \cite{H2CS_laser-Stark_1980} 
and $\nu _2$ \cite{H2CS_FIR_nu2_1993} in our first combined fit. Uncertainties of 0.001~cm$^{-1}$ were assigned 
to the laser Stark data \cite{H2CS_laser-Stark_1980}, 0.01~cm$^{-1}$ to the $\nu _4$ and $\nu _6$ data 
\cite{H2CS_D2CS_IR_re-est_1981} and 0.0005~cm$^{-1}$ initially to the $\nu _2$ data \cite{H2CS_FIR_nu2_1993} 
roughly based on the average residuals in the initial fits. The uncertainties of the $\nu _2$ data were reduced 
in the final fits to 0.00035~cm$^{-1}$. We obtained the $\nu _2$ data from the corresponding author as 
they were not available anymore as supplementary material through the journal.


\begin{table}
\tbl{Ground state spectroscopic parameters (MHz) of H$_2$CS from previous rotational study 
     (rot)\textsuperscript{a} and from combined fits with old (combined1)\textsuperscript{b} and new 
     (combined2)\textsuperscript{c} data of $\nu _4$, $\nu _6$ and $\nu _3$ along with H$_2$C$^{34}$S parameters.}
{
\begin{tabular}{lr@{}lr@{}lr@{}lr@{}l}
\toprule 
Parameter & \multicolumn{6}{c}{H$_2$CS} & \multicolumn{2}{c}{H$_2$C$^{34}$S} \\
\cmidrule{2-7}
          & \multicolumn{2}{c}{rot} & \multicolumn{2}{c}{combined1} & \multicolumn{2}{c}{combined2} &   \\
\midrule 
$A- (B+C)/2$             &  274437&.5932~(115)    &  274437&.5891~(113)    &  274437&.6073~(110)    &  274729&.46~(19)      \\
$(B+C)/2$                &   17175&.745955~(196)  &   17175&.746282~(147)  &   17175&.746270~(146)  &   16882&.911660~(111) \\
$(B-C)/4$                &     261&.6240523~(165) &     261&.6240459~(165) &     261&.6240453~(165) &     252&.793065~(73)  \\
$D_K \times 10^3$        &   23343&.78~(164)      &   23342&.99~(160)      &   23346&.11~(154)      &   23625&.~(58)        \\
$D_{JK} \times 10^3$     &     522&.2938~(43)     &     522&.3013~(40)     &     522&.2957~(40)     &     504&.8472~(48)    \\
$D_J \times 10^6$        &   19018&.75~(39)       &   19019&.24~(27)       &   19019&.28~(27)       &   18404&.267~(136)    \\
$d_1 \times 10^6$        & $-$1208&.429~(105)     & $-$1208&.557~(92)      & $-$1208&.534~(92)      & $-$1148&.572~(108)    \\
$d_2 \times 10^6$        &  $-$177&.3270~(222)    &  $-$177&.3113~(215)    &  $-$177&.3077~(214)    &  $-$165&.659~(120)    \\
$H_K \times 10^3$        &       5&.946~(35)      &       5&.914~(34)      &       5&.954~(33)      &       6&.00           \\
$H_{KJ} \times 10^6$     &   $-$28&.155~(86)      &   $-$28&.071~(83)      &   $-$28&.194~(82)      &   $-$28&.027~(106)    \\
$H_{JK} \times 10^6$     &       1&.50409~(270)   &       1&.50752~(236)   &       1&.50744~(235)   &       1&.41855~(70)   \\
$H_J \times 10^9$        &    $-$5&.81~(32)       &    $-$5&.51~(21)       &    $-$5&.50~(21)       &    $-$4&.913~(40)     \\
$h_1 \times 10^9$        &       3&.018~(141)     &       3&.222~(120)     &       3&.190~(120)     &       2&.765~(36)     \\
$h_2 \times 10^9$        &       1&.6472~(140)    &       1&.6374~(135)    &       1&.6373~(135)    &       1&.412~(49)     \\
$h_3 \times 10^9$        &       0&.3619~(73)     &       0&.3739~(68)     &       0&.3763~(68)     &       0&.3151~(99)    \\
$L_K \times 10^6$        &    $-$2&.109~(206)     &    $-$1&.903~(200)     &    $-$2&.056~(194)     &    $-$2&.09           \\
$L_{KKJ} \times 10^9$    &   $-$21&.36~(69)       &   $-$21&.69~(67)       &   $-$20&.84~(67)       &   $-$20&.92~(63)      \\
$L_{JK} \times 10^9$     &       0&.2032~(90)     &       0&.1855~(82)     &       0&.1883~(82)     &       0&.183          \\
$L_{JJK} \times 10^{12}$ &   $-$10&.32~(81)       &   $-$10&.66~(71)       &   $-$10&.73~(71)       &    $-$9&.4            \\
$L_J \times 10^{12}$     &       0&.833~(87)      &       0&.767~(58)      &       0&.766~(58)      &       0&.65           \\
$l_1 \times 10^{12}$     &    $-$0&.358~(47)      &    $-$0&.432~(40)      &    $-$0&.421~(40)      &    $-$0&.35           \\
$P_{KKJ} \times 10^{12}$ &   $-$18&.63~(180)      &   $-$18&.11~(177)      &   $-$20&.03~(175)      &   $-$19&.8            \\
\bottomrule 
\end{tabular}
}
\tabnote{
Notes: Watson's $S$ reduction was used in the representation $I^r$. Numbers in parentheses are one standard 
       deviation in units of the least significant figures. Parameters without uncertainties were estimated and 
       kept fixed in the analyses. See also Section~\ref{H2CS-results} for further details.\\
\textsuperscript{a}Ref.~\cite{H2CS_rot_2019} \\ 
\textsuperscript{b}Ground state data from Ref.~\cite{H2CS_rot_2019} and IR data from Refs.~\cite{H2CS_FIR_nu2_1993,H2CS_laser-Stark_1980,H2CS_D2CS_IR_re-est_1981}.\\ 
\textsuperscript{c}Ground state data from Ref.~\cite{H2CS_rot_2019} and IR data from Refs.~\cite{H2CS_FIR_nu2_1993,H2CS_IR_2008}. 
}
\label{tab_32ground-state-parameters}
\end{table}


We encountered labelling issues in particular in the laser Stark study \cite{H2CS_laser-Stark_1980}. 
The $\nu _3$ transitions should obey $a$-type selection rules with $\Delta K_a = 0$ and $\Delta K_c = \pm1$. 
They were mentioned as $a$-type transitions in the text but were frequently given as $x$-type transitions 
with $\Delta K_a = \Delta K_c = 0$ in the table. The applied Stark field and also the Lamb-dip technique, 
which was used for several transitions, may introduce intensities to otherwise forbidden transitions. 
We have attributed the assignments to allowed transitions wherever this appeared 
to be appropriate but retained assignments where this would cause too large residuals in the fit or 
if an allowed and a forbidden transition with similar quantum numbers were reported. It was necessary 
in other cases to modify $K_c$ or $K_a$ or both, also in the FTIR study of $\nu _4$ and $\nu _6$ 
\cite{H2CS_D2CS_IR_re-est_1981}, or to modify the vibrational assignments. Two $K_a = 1 - 0$ $Q$-branch 
transitions of $\nu _4$ \cite{H2CS_laser-Stark_1980} were reassigned from $J = 1$ and 2 to $J = 2$ and 4. 
Several own lines were omitted in the $\nu _2$ FTIR study. We included all but two in our fit and point out 
that some of the lines fitted better because they were treated as unresolved asymmetry doublets in our fit 
whereas only one line each was given in the initial line list. We omitted six and three transition frequencies 
from the laser Stark study \cite{H2CS_laser-Stark_1980} and from the $\nu _4$ and $\nu _6$ investigation 
\cite{H2CS_D2CS_IR_re-est_1981}, respectively, because of large residuals.

The parameter set was assembled starting from first-order Coriolis parameters from a quantum-chemical 
calculation. The values applied in intermediate fits were close to available values, but did not agree 
with any particular set. Therefore, the final values were taken from an MP2/aug-cc-pwCVQZ calculation 
as indicated in Section~\ref{sec_fitting}. Initial parameters $\Delta X$ were evaluated from previous 
work \cite{H2CS_laser-Stark_1980,H2CS_D2CS_IR_re-est_1981,H2CS_FIR_nu2_1993} as were estimates 
of the vibrational energies. The need for fitting any of the first-order Coriolis parameters 
was evaluated frequently. It became clear rather quickly that it was necessary to fit $G_a$(4,6) and 
$G_c$(3,6). The situation was more complex in the case of $G_b$(3,4). Trial fits in a late stage of 
the fitting process yielded values close to the initial one with reasonable uncertainties of $\sim$9~MHz, 
however, uncertainties not only of $\Delta (B+C)/2$ and $\Delta (B-C)/4$ of $v_3 = 1$ and $v_4 = 1$ increased 
substantially but through correlation also those of $\Delta (A- (B+C)/2)$ of $v_4 = 1$ and $v_6 = 1$ 
among several others and their values changed well outside the larger uncertainties. Therefore, $G_b$(3,4) 
was kept fixed to the initial value in the latest fits.

It turned out in the fitting process that the signs of the $G_i(x,y)$ cannot be changed independently 
of each other without significant deterioration of the quality of the fit. 
A sign change in $G_b$(2,4) afforded a sign change in $G_c$(2,6) if the sign of $G_a$(4,6) was retained; 
the same applied to $G_b$(3,4) and $G_c$(3,6). If, however, the sign of $G_a$(4,6) was changed it was 
necessary to change the signs of two other combinations of $G_b(x,y)$ or $G_c(x,y)$.

There appeared to be no dependence of the intensities of the rotational transitions upon sign change of 
the $G_i(x,y)$ among transitions having similar intensities as the observed ones. But in part substantial 
intensity modifications were calculated in the IR spectrum upon sign change of the $G_i(x,y)$. 
Applying the quantum-chemically calculated transition dipole moments as positive for four fundamental vibrations 
from Ref.~\cite{calc-int_H2CS_2020} slightly adjusted to take into account the experimental transition dipole 
moment ratios for the lower three fundamentals \cite{H2CS_IR_2008} it was fairly obvious from relative 
intensities in several $\nu _4$ and $\nu _6$ $Q$-branch transitions that $G_a$(4,6) has to be negative. 
The $P$-branch of $\nu _2$ appears to be slightly stronger than the $R$-branch according to Fig.~1 of 
Ref.~\cite{H2CS_FIR_nu2_1993}, affording $G_b$(2,4) and in turn $G_c$(2,6) to be positive. 
The effects of sign changes of $G_b$(3,4) and $G_c$(3,6) on the intensities are relatively similar 
but occur in a more crowded region of the spectrum. The choice of $G_b$(3,4) being negative and $G_c$(3,6) 
being positive appears to be in better agreement with the experimental spectrum \cite{H2CS_IR_2008} 
than the opposite sign choice.

Several $\Delta X$ were determined besides the vibrational energies for all four excited vibrational states. 
The selection differed among the states; most or all quartic $\Delta X$ were employed together with up to 
three of the seven sextic $\Delta X$. The parameters $\Delta (A- (B+C)/2)$ and $\Delta D_{JK}$ of $v_4 = 1$ 
and $v_6 = 1$ are of similar magnitude, but opposite sign. Even though they are comparatively small with 
respect to the corresponding ground state parameter omission of one caused a considerable deterioration of 
the quality of the fit that could not be easily accounted for otherwise. The $\Delta H_{KJ}$ for these states 
were comparatively large and only moderately well determined. Their values did not change significantly if only 
one of the two or both were employed in the fit. The values were constrained to be the same as a consequence. 
It was not possible to determine $\Delta D_K$ significantly for $v_6 = 1$ and for $v_3 = 1$. 
Several sextic distortion parameters were tested but many resulted in improvements that were largely negligible. 
The effect of $\Delta H_{J,6}$ was larger in some intermediate fits and more on the edge in the final fit.
Diverse sets of distortion corrections to the $G_i(x,y)$ were used in the fits, fairly large sets for $G_a$(4,6) 
and $G_c$(3,6) and no correction to $G_c$(2,6). There was no evidence for the need of any of the $F_{j,k}(x,y)$ 
in the fits.

The ground state rotational parameters from the final fit are given in Table~\ref{tab_32ground-state-parameters} 
together with those from our previous ground state study \cite{H2CS_rot_2019}, values from a second combined fit 
and values from a combined fit of H$_2$C$^{34}$S data. The $\Delta X_i$ from this fit are presented in 
Table~\ref{tab_32vib-state-parameters_old} while those of the second combined fit are gathered in 
Table~\ref{tab_32vib-state-parameters_new}. The interaction parameters from both fits as well as those from 
the H$_2$C$^{34}$S fit are finally given in Table~\ref{tab_interaction-parameters}.


\begin{table}
\tbl{Vibrational energies $E$ (cm$^{-1}$) and changes $\Delta X$ of spectroscopic parameters (MHz) of H$_2$CS 
     in excited vibrational states employing the old data of $\nu _4$, $\nu _6$ and $\nu _3$.}
{
\begin{tabular}{lr@{}lr@{}lr@{}lr@{}l}
\toprule 
Parameter & \multicolumn{2}{c}{$v_4 = 1$} & \multicolumn{2}{c}{$v_6 = 1$} & \multicolumn{2}{c}{$v_3 = 1$} & \multicolumn{2}{c}{$v_2 = 1$} \\
\midrule 
$E$                         &    990&.182542~(128)               &   991&.020175~(127)               &   1059&.204930~(123) &    1455&.495737~(25)   \\
$\Delta (A- (B+C)/2)$       & $-$143&.59~(49)                    &   632&.82~(40)                    & $-$222&.176~(281)    &    2510&.097~(113)     \\
$\Delta (B+C)/2$            &   $-$3&.198890~(229)               & $-$11&.07153~(50)                 & $-$108&.05638~(58)   &   $-$26&.976754~(233)  \\
$\Delta (B-C)/4$            &   $-$5&.03832~(158)                &    12&.52173~(167)                &   $-$0&.990805~(274) &       6&.931217~(282)  \\
$\Delta D_K \times 10^3$    &    666&.3~(264)                    &      &                            &       &              &    2041&.9~(26)        \\
$\Delta D_{JK} \times 10^3$ &   $-$4&.983~(284)                  &     2&.655~(126)                  &   $-$2&.641~(72)     &      23&.771~(287)     \\
$\Delta D_J \times 10^6$    &    190&.75~(59)                    &   105&.20~(46)                    &    158&.69~(44)      &  $-$105&.09~(59)       \\
$\Delta d_1 \times 10^6$    &  $-$11&.249~(296)                  & $-$85&.92~(37)                    &  $-$19&.914~(184)    &        &               \\
$\Delta d_2 \times 10^6$    &       &                            & $-$33&.259~(210)                  &   $-$6&.568~(72)     &        &               \\
$\Delta H_{KJ} \times 10^6$ &      7&.75~(39)\textsuperscript{a} &     7&.75~(39)\textsuperscript{a} &       &              &   $-$12&.64~(91)       \\
$\Delta H_J \times 10^9$    &       &                            &  $-$0&.436~(125)                  &      1&.811~(105)    &        &               \\
$\Delta h_1 \times 10^9$    &       &                            &     1&.594~(113)                  &       &              &        &               \\
\bottomrule 
\end{tabular}
}
\tabnote{
Notes: Watson's $S$ reduction was used in the representation $I^r$. $\Delta X = X\textsubscript{vib} - X\textsubscript{0}$. 
       Numbers in parentheses are one standard deviation in units of the least significant figures. Empty fields indicate 
       $\Delta X$ was not used in the final analysis.\\
\textsuperscript{a}Constrained to be equal, see Section~\ref{H2CS-results}.
}
\label{tab_32vib-state-parameters_old}
\end{table}


\begin{table}
\tbl{Vibrational energies $E$ (cm$^{-1}$) and changes $\Delta X$ of spectroscopic parameters (MHz) of H$_2$CS 
     in excited vibrational states employing the new data of $\nu _4$, $\nu _6$ and $\nu _3$.}
{
\begin{tabular}{lr@{}lr@{}lr@{}lr@{}l}
\toprule 
Parameter & \multicolumn{2}{c}{$v_4 = 1$} & \multicolumn{2}{c}{$v_6 = 1$} & \multicolumn{2}{c}{$v_3 = 1$} & \multicolumn{2}{c}{$v_2 = 1$} \\
\midrule 
$E$                         &    990&.182588~(19)                &   991&.020207~(19)                &   1059&.205183~(17)  &    1455&.495737~(25)   \\
$\Delta (A- (B+C)/2)$       & $-$142&.529~(245)                  &   634&.741~(242)                  & $-$222&.955~(106)    &    2510&.094~(113)     \\
$\Delta (B+C)/2$            &   $-$3&.198386~(213)               & $-$11&.07162~(49)                 & $-$108&.05631~(57)   &   $-$26&.977124~(226)  \\
$\Delta (B-C)/4$            &   $-$5&.04129~(147)                &    12&.51873~(156)                &   $-$0&.990983~(268) &       6&.931084~(280)  \\
$\Delta D_K \times 10^3$    &    771&.77~(213)                   &      &                            &  $-$26&.81~(230)     &    2041&.0~(26)        \\
$\Delta D_{JK} \times 10^3$ &   $-$5&.505~(265)                  &     2&.565~(123)                  &   $-$2&.604~(70)     &      24&.477~(260)     \\
$\Delta D_J \times 10^6$    &    190&.50~(58)                    &   105&.92~(46)                    &    157&.76~(44)      &  $-$105&.08~(59)       \\
$\Delta d_1 \times 10^6$    &  $-$11&.394~(295)                  & $-$85&.58~(36)                    &  $-$20&.321~(162)    &        &               \\
$\Delta d_2 \times 10^6$    &       &                            & $-$33&.150~(196)                  &   $-$6&.471~(69)     &        &               \\
$\Delta H_{KJ} \times 10^6$ &      8&.58~(38)\textsuperscript{a} &     8&.58~(38)\textsuperscript{a} &       &              &   $-$14&.07~(89)       \\
$\Delta H_J \times 10^9$    &       &                            &  $-$0&.452~(122)                  &      1&.748~(103)    &        &               \\
$\Delta h_1 \times 10^9$    &       &                            &     1&.540~(112)                  &       &              &        &               \\
\bottomrule 
\end{tabular}
}
\tabnote{
Notes: Watson's $S$ reduction was used in the representation $I^r$. $\Delta X = X\textsubscript{vib} - X\textsubscript{0}$. 
       Numbers in parentheses are one standard deviation in units of the least significant figures. Empty fields indicate 
       $\Delta X$ was not used in the final analysis.\\
\textsuperscript{a}Constrained to be equal, see Section~\ref{H2CS-results}.
}
\label{tab_32vib-state-parameters_new}
\end{table}


We became aware of an extensive high-resolution investigation of H$_2$CS in the 10$\umu$m region 
covering the $\nu _4$, $\nu _6$ and $\nu _3$ bands \cite{H2CS_IR_2008} at a relatively early stage 
of our own research and received a line list from one of the authors prior to publication. 
The interactions between the three excited vibrational states were taken into account 
as in earlier publications \cite{H2CS_laser-Stark_1980,H2CS_D2CS_IR_re-est_1981} but the effects 
of perturbations originating from $v_2 = 1$ were not considered even though a high-resolution study 
had been published in the meantime \cite{H2CS_FIR_nu2_1993}. We replaced the older $\nu _4$, $\nu _6$ 
and $\nu _3$ data with the new ones in order to evaluate the impact of these new data. We encountered 
massive labelling problems in that data set which have been discussed in a recent evaluation of experimental 
H$_2$CS data \cite{H2CS_marvel_2023}. Most mislabellings were encountered among the $\nu _4$ and $\nu _6$ 
at moderate and higher $K_a$ ($\gtrsim 2$). The vibrational identifier and the $K_c$ value had to be modified. 
It was necessary for some higher values of $J$ to decrease $K_a$ by 2 and raise $K_c$ by 2 compared 
with modifications at lower $J$. This was particularly peculiar in cases of unresolved asymmetry doublets 
in which one transition had the correct value of $K_a$ and one had a higher one by 2. There appear to 
be no labelling issues in the $\nu _3$ data. One of the transitions was reassigned and eight transition 
frequencies were omitted because of large residuals between reported frequencies and those calculated 
from the final fit. One line is a $K_a = 8$ $Q$-branch line in $\nu _3$, however, Table~1 of 
Ref.~\cite{H2CS_IR_2008} states $K_a \leq 7$ in $\nu _3$. Five omitted transition frequencies are 
$K_a = 10 - 9$ $R$-branch transition in $\nu _6$ correctly indicated in Table~1 of Ref.~\cite{H2CS_IR_2008} 
but not in the line list.

We assigned uncertainties of 0.0007~cm$^{-1}$ to the majority of the lines initially as suggested by 
the reported rms of the fit of exactly this value \cite{H2CS_IR_2008}. Some uncertainties were larger 
by a factor of 2 or 4 in accordance with indications in the line list. The uncertainties were reduced 
to 0.0005~cm$^{-1}$ and factors of 2 or 4 larger in the final fits. This data set required two 
additional parameters $\Delta D_{K,3}$ and $G_{a,KKK}$(4,6) besides the ones already employed 
to fit the old $\nu _4$, $\nu _6$ and $\nu _3$ data. We also tried to include these old IR data 
in the fit. While the data sets were compatible with each other the parameter values and their 
uncertainties changed insignificantly such that we omitted the old $\nu _4$, $\nu _6$ and $\nu _3$ 
data in the final second combined fit.

As indicated before, the ground state rotational parameters from this second combined fit are 
also in Table~\ref{tab_32ground-state-parameters}. The $\Delta X_i$ from this fit are given in 
Table~\ref{tab_32vib-state-parameters_new} and the interaction parameters are given in 
Table~\ref{tab_interaction-parameters}.

The new set of $\nu _4$, $\nu _6$ and $\nu _3$ data \cite{H2CS_IR_2008} consists of 3442 transitions 
corresponding to 2482 different lines with $J$/$K_a$ extending to 41/8, 35/8 and 34/7, respectively, 
and was fitted to 0.00048~cm$^{-1}$. The 599 $\nu _2$ transitions \cite{H2CS_FIR_nu2_1993} conform 
to 436 different lines with $J \leq 37$ and $K_a \leq 7$ and were reproduced in both fits to 
0.00035~cm$^{-1}$, also within the uncertainties on average. The 80 $\nu _4$, $\nu _6$ and $\nu _3$ 
transitions (70 lines) from a laser Stark study \cite{H2CS_laser-Stark_1980} with $J$/$K_a$ reaching 
13/3, 12/2 and 6/5, respectively, had an rms of 0.00105~cm$^{-1}$ and the 376 transitions (224 lines) 
from an older FTIR investigation of $\nu _4$ and $\nu _6$ \cite{H2CS_D2CS_IR_re-est_1981} extend to 
$J$/$K_a$ of 25/5 and 24/8, respectively, were reproduced to 0.0106~cm$^{-1}$; this is in both cases 
marginally above the attributed uncertainties. The 372 OSU rotational transitions (280 lines) cover 
$3 \leq J \leq 11$ and $K_a \leq 8$, 8, 7 and 6 for $v_4 = 1$, $v_6 = 1$, $v_3 = 1$ and $v_2 = 1$, 
respectively, with rms values of 50.7~kHz in the first combined fit and 51.4~kHz in the second combined fit, 
marginally above the assigned uncertainties. 
The measurements at the Universit\"at zu K\"oln resulted in 406 transition corresponding to 301 different 
lines with $16 \leq J \leq 41$ for $v_4 = 1$, $v_6 = 1$ and $v_3 = 1$ while $16 \leq J \leq 27$ in the 
case of $v_2 = 1$. The $K_a$ quantum numbers reached 12, 10, 11 and 9 for $v_4 = 1$, $v_6 = 1$, $v_3 = 1$ 
and $v_2 = 1$, respectively. The rms of 26.3~kHz in the first combined fit and 24.2~kHz in the second 
combined fit were on average slightly and marginally above the assigned uncertainties, respectively.


\begin{table}
\tbl{Interaction parameters (MHz) of H$_2$CS from from combined fits with old 
     (combined1)\textsuperscript{a} and new (combined2)\textsuperscript{b} data of 
     $\nu _4$, $\nu _6$ and $\nu _3$ along with H$_2$C$^{34}$S interaction parameters.}
{
\begin{tabular}{lr@{}lr@{}lr@{}l}
\toprule 
Parameter & \multicolumn{4}{c}{H$_2$CS} & \multicolumn{2}{c}{H$_2$C$^{34}$S} \\
\cmidrule{2-5}
          & \multicolumn{2}{c}{combined1} & \multicolumn{2}{c}{combined2} &   \\
\midrule 
                         & \multicolumn{6}{l}{$v_4 = 1$/$v_6 = 1$}                              \\
$G_a$                    & $-$300209&.33~(40)    & $-$300211&.424~(130)  & $-$299919&.0~(90)    \\
$G_{a,K}$                &        58&.371~(42)   &        58&.7267~(147) &        58&.76        \\
$G_{a,J} \times 10^3$    &       867&.1862~(271) &       867&.1433~(264) &       842&.702~(200) \\
$G_{a,KK} \times 10^3$   &     $-$29&.24~(88)    &     $-$31&.24~(38)    &     $-$31&.3         \\
$G_{a,JK} \times 10^6$   &       616&.8~(40)     &       615&.2~(39)     &       605&.          \\
$G_{a,KKK} \times 10^6$  &          &            &        37&.47~(288)   &        37&.6         \\
$G_{a,JKK} \times 10^9$  &       674&.3~(180)    &       658&.5~(175)    &       648&.          \\
$G_{a,JJK} \times 10^9$  &      $-$4&.82~(35)    &      $-$5&.32~(35)    &      $-$5&.2         \\
$G_{2a} \times 10^3$     &    $-$416&.07~(234)   &    $-$422&.20~(219)   &    $-$408&.          \\
$G_{4a} \times 10^6$     &      $-$7&.214~(177)  &      $-$7&.211~(171)  &      $-$6&.7         \\
                         &          &            &          &            &          &           \\
                         & \multicolumn{6}{l}{$v_2 = 1$/$v_4 = 1$}                              \\
$G_b$                    &     30660&.           &     30660&.           &     30118&.          \\
$G_{b,K}$                &        10&.196~(125)  &        10&.498~(113)  &        10&.33        \\
$G_{b,J} \times 10^3$    &     $-$65&.379~(262)  &     $-$65&.193~(259)  &     $-$62&.9         \\
$G_{2b} \times 10^3$     &        42&.271~(125)  &        42&.455~(120)  &        40&.3         \\
                         &          &            &          &            &          &           \\
                         & \multicolumn{6}{l}{$v_3 = 1$/$v_4 = 1$}                              \\
$G_b$                    &   $-$2177&.           &   $-$2177&.           &   $-$2172&.          \\
$G_{b,J} \times 10^3$    &     $-$97&.857~(56)   &     $-$97&.818~(56)   &     $-$95&.92        \\
$G_{b,JJ} \times 10^6$   &         1&.2886~(131) &         1&.2933~(127) &         1&.25        \\
$G_{2b} \times 10^3$     &         4&.823~(33)   &         4&.790~(31)   &         4&.59        \\
                         &          &            &          &            &          &           \\
                         & \multicolumn{6}{l}{$v_3 = 1$/$v_6 = 1$}                              \\
$G_c$                    &      9301&.155~(127)  &      9301&.148~(125)  &      9015&.270~(235) \\
$G_{c,K}$                &      $-$6&.3011~(259) &      $-$6&.3025~(256) &      $-$6&.373~(24)  \\
$G_{c,J} \times 10^3$    &     $-$45&.200~(75)   &     $-$45&.348~(75)   &     $-$43&.47        \\
$G_{c,KK} \times 10^3$   &        16&.310~(311)  &        16&.395~(306)  &        16&.02        \\
$G_{c,JK} \times 10^6$   &     $-$84&.42~(233)   &     $-$82&.33~(228)   &     $-$79&.          \\
$G_{c,JJ} \times 10^6$   &         0&.1212~(144) &         0&.1235~(132) &         0&.116       \\
$G_{2c} \times 10^3$     &        14&.050~(88)   &        13&.926~(72)   &        13&.12        \\
$G_{2c,K} \times 10^6$   &    $-$297&.5~(209)    &    $-$282&.1~(169)    &    $-$270&.          \\
$G_{2c,J} \times 10^9$   &    $-$470&.2~(140)    &    $-$441&.5~(125)    &    $-$420&.          \\
                         &          &            &          &            &          &           \\
                         & \multicolumn{6}{l}{$v_2 = 1$/$v_6 = 1$}                              \\
$G_c$                    &       459&.           &       459&.           &       480&.          \\
\bottomrule 
\end{tabular}
}
\tabnote{
Notes: Numbers in parentheses are one standard deviation in units of the least significant figures. Parameters values without 
       uncertainties were estimated and kept fixed in the analyses, see Section~\ref{results}. 
}
\label{tab_interaction-parameters}
\end{table}

\subsection{H$_2$C$^{34}$S}
\label{H2CS-34-results}

Having assigned rotational transition of H$_2$CS in the four lowest excited vibrational states up to 
fairly high values of $K_a$ we wondered if excited state transitions of H$_2$C$^{34}$S and even of H$_2^{13}$CS 
are assignable in the OSU spectral recordings. Quantum-chemical calculations were performed for both isotopologues 
to evaluate vibrational energies and first-order Coriolis coupling parameters $G_i(x,y)$. 
Improved ground state parameters were already available at that time although not to the extent as in our 
account on isotopic H$_2$CS \cite{H2CS_rot_2019}. Vibrational differences $\Delta X$ and centrifugal distortion 
corrections to the $G_i(x,y)$ were evaluated from the best available H$_2$CS values by scaling with isotopic 
ratios of appropriate powers of $A- (B+C)/2$, $(B+C)/2$ and $(B-C)/4$ in their ground vibrational states 
as done for various isotopic species in our previous study \cite{H2CS_rot_2019}.

Only some of the strongest $K_a = 0$ transitions in $v_4 = 1$ and $v_6 = 1$ were identified for H$_2^{13}$CS 
with some certainty and were not further considered. Quite extensive assignments could be made in the case of 
H$_2$C$^{34}$S with $3 \leq J \leq 11$ and $K_a$ extending to 6, 8, 6 and 5 for $v_4 = 1$, $v_6 = 1$, $v_3 = 1$ 
and $v_2 = 1$, respectively. Uncertainties of 100~kHz were applied throughout. No attempt was unfortunately made 
to record transitions of vibrationally excited H$_2$C$^{34}$S in Cologne.

We employed results of an MP2/aug-cc-pwCVQZ anharmonic force field calculation, as described in 
Section~\ref{sec_fitting}, to evaluate ground state rotational parameters, their vibrational changes, 
first-order Coriolis coupling parameters $G_i(x,y)$ and equilibrium quartic and sextic centrifugal 
distortion parameters of H$_2$C$^{34}$S. The experimental H$_2$CS values determined in 
Section~\ref{H2CS-results} were scaled with calculated H$_2$C$^{34}$S/H$_2$CS ratios wherever 
appropriate as was done for isotopic cyclopropenone \cite{c-H2C3O_rot_2021} and included the $G_i(x,y)$. 
To evaluate the distortion corrections to the $G_i(x,y)$, we scaled each of the H$_2$CS 
corrections with H$_2$C$^{34}$S/H$_2$CS ratios of the appropriate $G_i(x,y)$ and scaled these 
values with H$_2$C$^{34}$S/H$_2$CS ratios of the appropriate spectroscopic parameters; e.g., 
we scaled $G_{i,J}(x,y)$ with the ratio of $(B+C)/2$. We searched subsequently for parameters 
whose fitting would improve the fit as described in Section~\ref{sec_fitting}. 
Besides the obvious choices of $\Delta (B+C)/2$ and $\Delta (B-C)/4$ and some of the interaction parameters 
it became clear that fitting of one of the vibrational energies was necessary to achieve a satisfactory fit. 
The best result was obtained by fitting the energy of $v_4 = 1$. The ground state rotational 
data \cite{H2CS_rot_2019} were included in the final fit and the ground state parameters fitted as 
in the case of the H$_2$CS isotopologue. Fitting of further spectroscopic parameters, such as changes 
in the quartic centrifugal distortion parameters, had only minute effects on the quality of the fit 
and the resulting uncertainties or the changes in value were usually too large.


\begin{table}
\tbl{Vibrational energies $E$ (cm$^{-1}$) and changes $\Delta X$ of spectroscopic parameters (MHz) 
     of H$_2$C$^{34}$S in excited vibrational states.}
{
\begin{tabular}{lr@{}lr@{}lr@{}lr@{}l}
\toprule 
Parameter & \multicolumn{2}{c}{$v_4 = 1$} & \multicolumn{2}{c}{$v_6 = 1$} & \multicolumn{2}{c}{$v_3 = 1$} & \multicolumn{2}{c}{$v_2 = 1$} \\
\midrule 
$E$                         &    990&.02140~(11)  &   990&.355         &   1049&.828         &    1455&.391         \\
$\Delta (A- (B+C)/2)$       & $-$142&.68          &   635&.42          & $-$223&.19          &    2512&.78          \\
$\Delta (B+C)/2$            &   $-$3&.25192~(173) & $-$11&.25386~(193) & $-$105&.36274~(121) &   $-$26&.27430~(115) \\
$\Delta (B-C)/4$            &   $-$4&.8659~(50)   &    12&.1687~(49)   &   $-$0&.98868~(171) &       6&.71131~(180) \\
$\Delta D_K \times 10^3$    &    776&.1           &      &             &  $-$27&.0           &    2053&.0           \\
$\Delta D_{JK} \times 10^3$ &   $-$5&.321         &     2&.48          &   $-$2&.52          &      23&.66          \\
$\Delta D_J \times 10^6$    &    184&.3           &   102&.5           &    152&.6           &  $-$101&.6           \\
$\Delta d_1 \times 10^6$    &  $-$10&.83          & $-$81&.3           &  $-$19&.3           &        &             \\
$\Delta d_2 \times 10^6$    &       &             & $-$30&.96          &   $-$6&.04          &        &             \\
$\Delta H_{KJ} \times 10^6$ &      8&.55          &     8&.55          &       &             &   $-$14&.0           \\
$\Delta H_J \times 10^9$    &       &             &  $-$0&.4           &      1&.5           &        &             \\
$\Delta h_1 \times 10^9$    &       &             &     1&.33          &       &             &        &             \\
\bottomrule 
\end{tabular}
}
\tabnote{
Notes: Watson's $S$ reduction was used in the representation $I^r$. $\Delta X = X\textsubscript{vib} - X\textsubscript{0}$. 
       Numbers in parentheses are one standard deviation in units of the least significant figures. Parameters values without 
       uncertainties were estimated and kept fixed in the analyses. Empty fields indicate  $\Delta X$ was not used in the final analysis.
}
\label{tab_34vib-state-parameters}
\end{table}


The resulting ground state spectroscopic parameters are presented in Table~\ref{tab_32ground-state-parameters} 
as well, the interaction parameters are given in Table~\ref{tab_interaction-parameters} and the vibrational 
energies and changes $\Delta X$ of spectroscopic parameters can be found in Table~\ref{tab_34vib-state-parameters}. 
The excited state data alone were fitted to 81.4~kHz on average, marginally improving to 79.1~kHz after addition 
of the ground state rotational data.

\subsection{Thioketene}
\label{H2C2S-results}

A peculiar series of lines in the vicinity of the H$_2$C$^{34}$S lines remained unassigned. 
The appearance in the Loomis$-$Wood diagrams resembled somewhat those of thioformaldehyde 
even though it is even closer to the prolate symmetric top limit with a $\kappa$ of $-$0.9992. 
However, it became evident quite quickly that between two seemingly adjacent lines there were 
two more lines of similar intensity and regularly spaced. The molecule had to be a heavier one 
with $(B+C)/2$ nearly one-third of the H$_2$C$^{34}$S value. Thioketene was a plausible candidate. 
The assignment could be established with the help of previous laboratory data 
\cite{H2CCS_rot_1979a,H2CCS_rot_1979b,H2CCS_rot_1980}. The previous data extended to $J = 20 - 19$ 
up to 226~GHz. We extended the line list from 234 to 361~GHz corresponding to $20 \leq J \leq 32$ 
with $K_a \leq 7$.

Uncertainties of 20~kHz were assigned to the $J = 3 - 2$ data \cite{H2CCS_rot_1979a}, 10~kHz to 
a set of $K_a = 1$ $Q$-branch transitions \cite{H2CCS_rot_1979b}, 15~kHz to the remaining extensive 
microwave and millimetre wave data \cite{H2CCS_rot_1980} and initially 50~kHz to the OSU data which 
were reduced to 30~kHz in the last fit.


\begin{table}
\tbl{Ground state spectroscopic parameters (MHz) of H$_2$C$_2$S from present data set in comparison 
     to values from a high-resolution IR study and from a rotational study.}
{
\begin{tabular}{lr@{}lr@{}lr@{}l}
\toprule 
Parameter & \multicolumn{2}{c}{present} & \multicolumn{4}{c}{previous} \\
\cmidrule{4-7}
                      &       &               & \multicolumn{2}{c}{Ref.~\cite{H2CCS_IR_1996}} 
 & \multicolumn{2}{c}{Ref.~\cite{H2CCS_rot_1980}\textsuperscript{a}} \\
\midrule 
$A- (B+C)/2$          & 280916&.1~(227)      & 280834&.764~(78)     & 281083&.~(78)        \\
$(B+C)/2$             &   5601&.994368~(127) &   5601&.993823~(105) &   5601&.994204~(218) \\
$(B-C)/4$             &     28&.740720~(18)  &     28&.740742~(17)  &     28&.74086~(32)   \\
$D_K            $     &     24&.6191         &     24&.6191~(30)    &     24&.6191         \\
$D_{JK} \times 10^3$  &    168&.330~(46)     &    168&.330~(26)     &    168&.212~(74)     \\
$D_J \times 10^3$     &      1&.08513~(10)   &      1&.08418~(16)   &      1&.08559~(36)   \\
$d_1 \times 10^6$     &  $-$25&.426~(34)     &  $-$25&.470~(33)     &  $-$25&.55~(63)      \\
$d_2 \times 10^6$     &   $-$5&.908~(91)     &   $-$6&.140~(75)     &   $-$5&.15~(32)      \\
$H_K \times 10^3$     &      7&.387          &      7&.387~(33)     &      7&.387          \\
$H_{KJ} \times 10^6$  & $-$400&.0~(44)       & $-$414&.4~(21)       & $-$413&.5~(70)       \\
$H_{JK} \times 10^9$  &    617&.3~(130)      &    582&.8~(113)      &    714&.7~(184)      \\
$L_{KKJ} \times 10^9$ & $-$866&.~(144)       & $-$338&.~(61)        & $-$517&.~(225)       \\
$L_{JK} \times 10^9$  &   $-$2&.365~(272)    &   $-$2&.73~(52)      &       &              \\
$P_{KKJ} \times 10^9$ &  $-$52&.36~(148)     &  $-$58&.04~(60)      &  $-$54&.61~(229)     \\
$P_{KJ} \times 10^9$  &   $-$0&.109          &   $-$0&.109~(7)      &       &              \\
\bottomrule 
\end{tabular}
}
\tabnote{
Notes: Watson's $S$ reduction was used in the representation $I^r$. Numbers in parentheses are one standard 
       deviation in units of the least significant figures. Parameters without uncertainties were taken from 
       Ref.~\cite{H2CCS_IR_1996} and kept fixed in the analyses. Empty fields indicate parameters 
       not used in the analysis.\\
\textsuperscript{a}Refit of Ref.~\cite{H2CCS_rot_1980}, see Section~\ref{H2C2S-results}.  
}
\label{tab_H2C2S-parameters}
\end{table}


The resulting spectroscopic parameters are provided in Table~\ref{tab_H2C2S-parameters} together with 
parameters from a far- and mid-IR spectroscopic study \cite{H2CCS_IR_1996} which also took into 
account previous rotational data and those of a refit of the data from Ref.~\cite{H2CCS_rot_1980}. 
The refit permits all parameters used in the fits to be compared directly in values and in uncertainties. 
The rms values are respectively 12.9~kHz \cite{H2CCS_rot_1979a}, 8.9~kHz \cite{H2CCS_rot_1979b}, 
15.9~kHz \cite{H2CCS_rot_1980} and finally 26.5~kHz for the OSU data.

\section{Discussion}
\label{discussion}

An extensive set of spectroscopic parameters of eighth-order plus one decic distortion parameter were 
required to fit the ground state rotational data of H$_2$CS shown in Table~\ref{tab_32ground-state-parameters} 
whereas only parameters up to sixth order were needed in the diagonal part of the Hamiltonian to fit the excited state 
data of this isotopologue, see Tables~\ref{tab_32vib-state-parameters_old} and \ref{tab_32vib-state-parameters_new}. 
Explanations are certainly that no $\Delta K_a = 0$ $Q$-branch transitions and no transitions involving 
$\Delta K_a = 2$ are in the fit for the excited states because these lines are weaker than corresponding 
ground state transitions by factors of around 100 and more. In addition, the coverage of $R$-branch transitions 
is also less extensive in particular in the Cologne data. 
Some may question the presence of parameters in the fit such as $\Delta H_{J,6}$. 
It is not surprising that $F_{j,k}(x,y)$ were not required in the fit because H$_2$CS is rather close to the 
prolate symmetric limit and for a symmetric top rotor either $G_i(x,y)$ or $F_{j,k}(x,y)$ may be used to fit a 
particular Coriolis or related interaction.

The successful fit of essentially all $\nu _2$ transition frequencies \cite{H2CS_FIR_nu2_1993} at a presumably 
improved rms and the better fit of the $\nu _4$, $\nu _6$ and $\nu _3$ data set from Ref.~\cite{H2CS_IR_2008} 
are testament to the importance of including $v_ 2 = 1$ and its perturbation into the analysis. 
There are a small number of rotational transitions with relatively large residuals, for example two higher $J$ 
transitions of $v_3 = 1$ with $K_a = 11$. It is frequently difficult to determine if such residuals at the edge 
of the data set are a result of missing parameters that cannot be determined yet with enough significance or 
if the lines have been misjudged. The slight increase from one $J$ to the next leaves room for the first 
interpretation but is statistically not meaningful.

A comparison of the uncertainties in the excited state parameters of the first combined fit in 
Table~\ref{tab_32vib-state-parameters_old} with the ones of the second combined fit in 
Table~\ref{tab_32vib-state-parameters_new} is instructive. The uncertainties of parameters with $J$ dependence 
hardly change for $v_4 = 1$, $v_6 = 1$ and $v_3 =1$ suggesting their uncertainties are predominantly determined 
through the rotational data. This is understandable as the rotational data are more accurate and extend to 
similar or mostly even higher quantum numbers. The larger number of rovibrational transition frequencies 
moderates this aspect, but only somewhat. The situation is different for the vibrational energies and the 
purely $K$-dependent parameters despite the use of one diagonal and one off-diagonal parameter more in the fit. 
This indicates that their uncertainties are mostly established through the rovibrational data. In the absence 
of $\Delta K_a = 2$ transitions in the excited states data the rotational data still contribute more indirectly 
to the purely $K$-dependent parameters through improving the accuracies of $J$-dependent parameters and through 
sampling resonances. The sampling of resonances may also contribute directly to establishing the $K$-level 
structure even of a symmetric top rotor if transitions within each state at the resonance are observed along 
with perturbation mediated transitions between the states as in multiple cases in the excited state spectra 
of methyl cyanide \cite{MeCN_up2v4eq1_etc_2021}. The effect of the rovibrational data on the uncertainties 
in the ground vibrational states in Table~\ref{tab_32ground-state-parameters} is much less pronounced because 
of $\Delta K_a = 2$ transitions from rotational data \cite{H2CS_rot_2019} and from ground state combination 
differences extracted from an electronic spectrum \cite{H2CS_A-X_1994}. The rovibrational data were also 
important for $G_a$(4,6) and its purely $K$-dependent distortion corrections in Table~\ref{tab_interaction-parameters}. 
Remarkable are also some improvements in higher order corrections to $G_c$(3,6).

We compare next selected spectroscopic parameters from our study with those from other experimental studies and from 
quantum-chemical calculations. Coriolis or other rotational resonances do not affect vibrational energies. 
Since the experimental rovibrational spectra access transitions with low $J$ and $K_a$ quantum numbers differences 
in the fitting should result in minute differences in the vibrational energies at most.


\begin{table}
\tbl{Selected interaction parameters (MHz) of H$_2$CS from from the second combined fit of this work (TW) 
     in comparison to values from previous experimental works and to values from quantum-chemical calculations (QC).} 
{
\begin{tabular}{lr@{}lr@{}lr@{}lr@{}llr@{}lr@{}l}
\toprule 
Parameter & \multicolumn{8}{c}{Experimental} & & \multicolumn{4}{c}{QC} \\
\cmidrule{2-9} \cmidrule{11-14} 
          & \multicolumn{2}{c}{TW} & \multicolumn{2}{c}{Ref.~\cite{H2CS_IR_2008}\textsuperscript{a}} & 
          \multicolumn{2}{c}{Ref.~\cite{H2CS_D2CS_IR_re-est_1981}} &  \multicolumn{2}{c}{Ref.~\cite{H2CS_laser-Stark_1980}} & 
          & \multicolumn{2}{c}{CC\textsuperscript{b}} & \multicolumn{2}{c}{MP2\textsuperscript{c}} \\
\midrule 
$G_a(4,6) \times 10^{-3}$  & $-$300&.2114                   & $-$300&.37 &   300&.228 & 299&.918 & &  299&.33 &  299&.67 \\
$F_{bc}(4,6)$              &       &                        &       &    &    58&.    &  49&.    & &     &    &     &    \\
$G_{a,K}(4,6)$             &     58&.37                     &  $-$77&.7  & $-$56&.0   &    &     & &     &    &     &    \\
$G_{a,J}(4,6)$             &      0&.86719                  &   $-$0&.55 &  $-$2&.6   &    &     & &     &    &     &    \\
$G_{2a}(4,6)$              &   $-$0&.416                    &       &    &     2&.    &    &     & &     &    &     &    \\
$G_b(3,4) \times 10^{-3}$  &   $-$2&.177\textsuperscript{d} &      4&.55 &     4&.9   &   1&.6   & & $-$2&.44 & $-$2&.18 \\
$G_{b,K}(3,4)$             &       &                        &     45&.   &      &     &    &     & &     &    &     &    \\
$G_{b,J}(3,4) \times 10^3$ &  $-$97&.86                     &    114&.   &      &     &    &     & &     &    &     &    \\
$G_c(3,6) \times 10^{-3}$  &      9&.3012                   &      2&.30 &    12&.2   &   9&.2   & &    8&.73 &    8&.93 \\
$G_{c,K}(3,6)$             &   $-$6&.30                     &  $-$16&.   &      &     &    &     & &     &    &     &    \\
$G_{c,J}(3,6) \times 10^3$ &  $-$45&.20                     & $-$343&.   &      &     &    &     & &     &    &     &    \\
\bottomrule 
\end{tabular}
}
\tabnote{
Notes: The last digit of each value is uncertain. Empty fields indicate values not determined. 
       The signs of the $G_i(x,y)$ are usually not determinable in the fits but matter for the IR intensities 
       in the present case, see also Section~\ref{H2CS-results}.\\
\textsuperscript{a}A value of $\sim$0.004~kHz was also determined for $G_{a,JJ}(4,6)$.\\
\textsuperscript{b}Coupled cluster calculation CCSD(T)/cc-pVTZ from Ref.~\cite{H2CS_FF_ai_1994}.\\
\textsuperscript{c}MP2/aug-cc-pwCVQZ from this work, see Section~\ref{sec_fitting}.\\
\textsuperscript{d}Kept fixed in the analysis, see Section~\ref{H2CS-results}.
}
\label{tab_comp_interaction}
\end{table}


A subset of interaction parameters of H$_2$CS are compared in Table~\ref{tab_comp_interaction}. 
We discuss the signs of the $G_i(x,y)$ separately as these are usually not determinable from the fits. 
The $G_a(4,6)$ values all agree well or very well in magnitude. The agreement is also quite good 
among the two quantum-chemical values of $G_b(3,4)$ and among the present $G_c(3,6)$ and the quantum-chemical values. 
The agreement is quite different between these experimental $G_i(x,y)$ and our values as well as 
for the distortion correction as far as they were determined. 
It is noteworthy that the corrections to $G_a(4,6)$ from the more recent IR study \cite{H2CS_IR_2008} are similar 
in magnitude to ours but of opposite sign. It is difficult to draw conclusions from this aspect as the agreement 
is worse for the interaction parameters involving $v_3 = 1$. A plausible explanation for the deviations are 
perturbations from $v_2 = 1$ which were not taken into account. In fact, $F_{bc}(4,6)$ was introduced in the 
two early IR studies \cite{H2CS_laser-Stark_1980,H2CS_D2CS_IR_re-est_1981} to account for perturbations of 
$v_4 = 1$ and $v_6 = 1$ by $v_2 = 1$ as specifically mentioned in the laser Stark study \cite{H2CS_laser-Stark_1980}. 
We introduced parameters into the fit to model interactions of $v_2 = 1$ with $v_4 = 1$ and $v_6 = 1$, so 
the absence of $F_{bc}(4,6)$ in our fit does not argue against that proposition. One of the early IR studies 
\cite{H2CS_D2CS_IR_re-est_1981} employed also distortion corrections to $G_a(4,6)$ of which $G_{a,K}(4,6)$ agrees 
well whereas there are factors of a few in the other two cases.


\begin{table}
\tbl{Selected low order $\Delta X_i$ parameters (MHz) of H$_2$CS from from the second combined fit of this work (TW) 
     in comparison to values from previous experimental works and to values from quantum-chemical calculations (QC).} 
{
\begin{tabular}{lr@{}lr@{}lr@{}lr@{}lr@{}llr@{}l}
\toprule 
Parameter & \multicolumn{10}{c}{Experimental} & & \multicolumn{2}{c}{QC} \\
\cmidrule{2-11} \cmidrule{13-14}
          & \multicolumn{2}{c}{TW} & \multicolumn{2}{c}{Ref.~\cite{H2CS_IR_2008}} & \multicolumn{2}{c}{Ref.~\cite{H2CS_FIR_nu2_1993}} & 
          \multicolumn{2}{c}{Ref.~\cite{H2CS_D2CS_IR_re-est_1981}} & \multicolumn{2}{c}{Ref.~\cite{H2CS_laser-Stark_1980}} & 
          & \multicolumn{2}{c}{CC\textsuperscript{a}} \\
\midrule 
$\Delta A_4 \times 10^{-3}$         &   $-$0&.1457 &      3&.65  &      &      &       &     &   $-$2&.17                 & &   $-$4&.143 \\
$\Delta B_4$                        &  $-$13&.281  &  $-$71&.0   &      &      &  $-$68&.    & $-$155&.                   & &  $-$73&.    \\
$\Delta C_4$                        &      6&.884  &      5&.6   &      &      &     15&.    &     32&.                   & &      4&.    \\
$\Delta D_{K,4}$                    &      0&.772  &   $-$0&.9   &      &      &       &     &       &\textsuperscript{b} & &       &     \\
$\Delta A_6 \times 10^{-3}$         &      0&.6237 &   $-$3&.18  &      &      &       &     &      2&.43                 & &      4&.817 \\
$\Delta B_6$                        &     13&.966  &     18&.3   &      &      &  $-$87&.    &  $-$84&.                   & &     12&.    \\
$\Delta C_6$                        &  $-$36&.11   &  $-$75&.7   &      &      &     91&.    &    159&.                   & &  $-$40&.    \\
$\Delta D_{K,6}$                    &       &      &      1&.7   &      &      &       &     &       &\textsuperscript{b} & &       &     \\
$\Delta (A_4 + A_6) \times 10^{-3}$ &      0&.4779 &      0&.479 &      &      &      0&.228 &      0&.17                 & &      0&.574 \\
$\Delta D_{K,4} + \Delta D_{K,6}$   &      0&.772  &      0&.73  &      &      &       &     &       &\textsuperscript{b} & &       &     \\
$\Delta A_3 \times 10^{-3}$         &   $-$0&.3110 &   $-$0&.330 &      &      &   $-$0&.329 &   $-$0&.329                & &   $-$0&.332 \\
$\Delta B_3$                        & $-$110&.0383 & $-$115&.    &      &      & $-$120&.    &  $-$84&.                   & & $-$109&.    \\
$\Delta C_3$                        & $-$106&.074  &  $-$69&.7   &      &      & $-$140&.    & $-$102&.                   & & $-$102&.    \\
$\Delta D_{K,3}$                    &      0&.027  &      0&.2   &      &      &       &     &       &\textsuperscript{b} & &       &     \\
$\Delta A_2 \times 10^{-3}$         &      2&.4831 &       &     &     2&.4763 &       &     &       &                    & &      2&.898 \\
$\Delta B_2$                        &  $-$13&.1150 &       &     &    54&.31   &       &     &       &                    & &     46&.    \\
$\Delta C_2$                        &  $-$40&.9393 &       &     & $-$39&.49   &       &     &       &                    & &  $-$41&.    \\
$\Delta D_{K,2}$                    &      2&.041  &       &     &     1&.8    &       &     &       &                    & &       &     \\
\bottomrule 
\end{tabular}
}
\tabnote{
Notes: The last digit of each value is uncertain, except for $\Delta B_3$ and $\Delta C_3$ from Ref.~\cite{H2CS_D2CS_IR_re-est_1981} where it is one digit earlier. 
       Empty fields indicate usually values not determined. Please note that some studies \cite{H2CS_IR_2008,H2CS_laser-Stark_1980} 
       employed the A reduction only but differences are small with respect to the uncertainties and the quoted digits.\\
\textsuperscript{a}Coupled cluster calculation CCSD(T)/cc-pVTZ from Ref.~\cite{H2CS_FF_ai_1994}.\\
\textsuperscript{b}$\Delta K$ values are given, but are very uncertain.
}
\label{tab_comp_Deltas}
\end{table}


The early IR studies \cite{H2CS_laser-Stark_1980,H2CS_D2CS_IR_re-est_1981} did not address the signs of the 
$G_i(x,y)$. The later IR study \cite{H2CS_IR_2008} presents $G_a(4,6)$ as negative and $G_b(3,4)$ as well as 
$G_c(3,6)$ as positive. But it is not clear if the signs from that work can be compared directly because only 
the operator associated with $G_c(3,6)$ was defined as imaginary whereas all $G_i(x,y)$ operators are commonly 
defined as imaginary. With $G_a(4,6)$ being negative it was necessary in our fits that the signs of $G_b(3,4)$ 
and $G_c(3,6)$ differ. The $P$-branch of $\nu _2$ was calculated to be stronger than the $R$-branch in 
Ref.~\cite{H2CS_ai_egy-etc_2013}, in accordance with our interpretation of Fig.~1 of Ref.~\cite{H2CS_FIR_nu2_1993}, 
but in contrast to Ref.~\cite{H2CS_ExoMol_2023}.

Some low order $\Delta X_i$ parameters of H$_2$CS are shown for comparison purpose in Table~\ref{tab_comp_Deltas}. 
The agreement is often quite poor for the experimental $v_4 = 1$ and $v_6 = 1$ parameters. This can be attributed 
to the $a$-type Coriolis resonance, its treatment and the data coverage in the case of $\Delta A$ and $\Delta D_K$. 
Their sums $\Delta A_4 + \Delta A_6$ and $\Delta D_{K,4} + \Delta D_{K,6}$ are better constrained and agree much better. 
The agreement is very good and good for these values from Ref.~\cite{H2CS_IR_2008} even though the individual 
parameters differ very much. The sum $\Delta A_4 + \Delta A_6$ was only determined in Ref.~\cite{H2CS_D2CS_IR_re-est_1981}. 
This is an indication of the importance of the rotational data from this study for constraining the spectroscopic parameters. 
And the uncertainty of $\Delta A_4 + \Delta A_6$ is even in the present study with 119~kHz much smaller than that 
of $\Delta A_4 - \Delta A_6$ with 472~kHz. Correlations with other parameters may complicate the situation in particular 
for previous studies; constraints in the present study, including the omission of some parameter differences, 
have helped to reduce correlation.

The $v_3 = 1$ $\Delta X_i$ from Refs.~\cite{H2CS_laser-Stark_1980,H2CS_D2CS_IR_re-est_1981} agree reasonably well 
with the present values if we take into account the very small $\nu _3$ data set in both studies. Therefore, we 
attribute the disagreement of the $v_4 = 1$ and $v_6 = 1$ values of $\Delta B$ and $\Delta C$ in these studies to 
the use of $F_{bc}(4,6)$ in their fits. The disagreement of $\Delta C_3$ and $\Delta C_6$ from Ref.~\cite{H2CS_IR_2008} 
with our values is connected with the very different $G_c(3,6)$ value.

The $\Delta X_2$ from Ref.~\cite{H2CS_FIR_nu2_1993} agree quite well with ours except for $\Delta B_2$ which is 
clearly a consequence of neglecting the distant but strong Coriolis interaction with $v_4 = 1$. This is fairly 
well mirrored in the deviation of $\Delta B_4$ from Ref.~\cite{H2CS_IR_2008} and even better mirrored in the 
$\Delta B_4$ and $\Delta B_2$ values from a quantum-chemical study \cite{H2CS_FF_ai_1994}. It is necessary 
to mention that quantum-chemical force field calculations usually do not determine resonance effects on 
spectroscopic parameters except possibly for the impact of Fermi or other anharmonic resonances on the 
vibrational energies. Moreover, quantum-chemical calculations provide commonly first-order vibrational 
corrections $\alpha _j^{B^i}$ associated with the $i$-axis rotational parameter and the vibration $\omega _j$ 
and not the $\Delta X^i_j$. The first-order vibrational corrections $\alpha _j^{B^i}$ are defined via 
\[ 
B^i_v = B^i_e - \sum_{j} \alpha _j^{B^i} (v_j + \frac{1}{2})   + \sum_{j \le k} \gamma _{jk}^{B^i} (v_j + \frac{1}{2})(v_k + \frac{1}{2}) + ...
\] 
where $B^i_e$ is the equilibrium rotational parameter associated with the $i$-axis, $B^i_v$ is the rotational 
parameter in an excited state $v$, $v_j$ and $v_k$ are the excitation quanta of $\nu _j$ and $\nu _k$ in 
this state $v$ and the $\gamma _{jk}^{B^i}$ are second-order vibrational corrections; higher order corrections 
may be defined equivalently. The equation $\Delta X^i_j = -\alpha _j^{B^i}/2$ is commonly assumed 
but holds only strictly if the $\gamma _{jk}^{B^i}$ and so on were zero. The $\Delta X^i_j$ taken from 
a quantum-chemical calculation \cite{H2CS_FF_ai_1994} agree well to reasonably well if the values are not 
or not strongly affected by resonances. These include $\Delta C_4$, $\Delta B_6$, $\Delta B_3$, $\Delta C_2$ 
and to a lesser extend $\Delta A_4$ + $\Delta A_6$, $\Delta A_3$ and $\Delta A_2$. The deviations are also 
quite small for $\Delta C_6$ and $\Delta C_3$ but would probably agree even better with our values if 
the effect of $G_c(3,6)$ is taken into account.

We have inspected calculated transition frequencies of H$_2$CS \cite{H2CS_ExoMol_2023} derived from a 
quantum-chemically calculated potential energy surface that was refined taking evaluated experimental data 
\cite{H2CS_marvel_2023} into account. We inspected $J = 3 - 2$ transition frequencies of the five lowest 
vibrational states and compared these with values derived from our second combined analysis. The agreements 
were good (deviations much less than 100~kHz) to reasonable (exceeding about 1~MHz). The largest deviations 
for these low-$J$ transitions were $\sim$1.0~MHz in $v_4 = 1$ and $v_6 = 1$, about 1.6~MHz in $v = 0$, 
$\sim$2.0~MHz in $v_3 = 1$ and almost 4~MHz in $v_2 = 1$. Further refinement may be possible taking our 
present H$_2$CS transition frequencies into account.

The data of H$_2$C$^{34}$S in excited vibrational states could be fitted well with varying only a relatively 
small number of spectroscopic parameters in Table~\ref{tab_34vib-state-parameters}. Unsurprisingly, these 
involved changes associated with $B$ and $C$ and a rather small number of interaction parameters, see 
Table~\ref{tab_interaction-parameters}. 
The change in $\Delta (B+C)/2$ from the initial values is $\sim$0.1~MHz for $v_4 = 1$ up to $\sim$0.9~MHz 
for $v_6 = 1$. The change in $\Delta (B-C)/4$ is $\sim$0.07~MHz in the case of $v_6 = 1$ and much smaller 
otherwise. It is difficult to evaluate how different the $\Delta (A- (B+C)/2)$ may be; deviations of some 
megahertz cannot be ruled out. No corrections to quartic distortion parameters needed to be released 
which indicates the quality of the applied scaling procedure \cite{c-H2C3O_rot_2021}. 
The determination of one vibrational energy ($v_4 = 1$) through the Coriolis perturbations in the spectrum 
improved the quality of the fit substantially. The change from the initial estimate of 990.017~cm$^{-1}$ 
is very small but should be viewed with some caution nevertheless because its value will depend somewhat 
on the exact energies of the remaining states and on the exact values of the $\Delta (A- (B+C)/2)$. 
Improvements in the ground state spectroscopic parameters of H$_2$C$^{34}$S in 
Table~\ref{tab_32ground-state-parameters} with respect to earlier values \cite{H2CS_rot_2019} should also 
be viewed with caution in particular for the purely $K$-dependent parameters. The application of realistic 
uncertainties of the $\Delta (A- (B+C)/2)$ may eliminate these improvements. The assigned uncertainties 
of 100~kHz to the experimental transition frequencies are quite appropriate at an rms of $\sim$80~kHz.

The additional transition frequencies of H$_2$C$_2$S lead to improvements in the spectroscopic parameters 
that are between slight and substantial with respect to the previous rotational study \cite{H2CCS_rot_1980}. 
The change in values of $A- (B+C)/2$ and $d_2$ may be connected but may also be a result of the two more 
spectroscopic parameters of which one was kept fixed in the analysis. The comparison is less favourable 
with respect to values from Ref.~\cite{H2CCS_IR_1996} where $D_K$ and $H_K$ were fitted and still most 
parameters display somewhat smaller uncertainties. Our $L_{JK}$ value has a smaller uncertainty but that 
is merely a consequence of fixing $P_{KJ}$ in our fit because it was not determinable with sufficient 
significance. A combined fit of our new data with those employed in Ref.~\cite{H2CCS_IR_1996} could 
result in a further improvement of the thioketene ground state parameters.

It is possible that transitions of thioketene with $K_a = 8$ or 9 are present in our OSU recordings 
but it would be very difficult to locate them as $v = 0$ is perturbed by $v_9 = 1$ and $v_9 = 1$ is 
part of a massive resonance system \cite{H2CCS_IR_1996}. The $v = 0$ energy of $K_a = 11$ is below 
$K_a = 10$ of $v_9 = 1$ whereas it is calculated to be opposite for the next higher values of $K_a$. 
Another detail not discussed in that work is that $K_a = 7$ of $v = 0$ is very close to 
$K_a = 4$ of $v_9 = 1$ which may not matter for data recorded with IR accuracy but may well 
affect the rotational data obtained with microwave accuracy. 
We mention also that Ref.~\cite{H2CCS_IR_1996} presented a combined analysis of $v = 0$, $v_9 = 1$ 
and $v_6 = 1$ in which the perturbation of $v = 0$ by $v_9 = 1$ was taken into account.

\section{Conclusion}
\label{conclusion}

We have presented assignments of rotational transitions pertaining to H$_2$CS and H$_2$C$^{34}$S in 
their lowest four excited vibrational states along with analyses of their intricate Coriolis perturbations. 
Sets of IR data were employed in the case of the main isotopologue. The results highlight the importance 
of taking resonances with $v_2 = 1$ into account in the analyses of the three lowest fundamental states 
of the main isotopologue which has been carried out here for the first time. Not only was it possible 
to fit essentially all $\nu _2$ data well, in contrast to the initial study \cite{H2CS_FIR_nu2_1993}, 
but also to achieve a better fit of an extensive set of $\nu _4$, $\nu _6$ and $\nu _3$ data 
\cite{H2CS_IR_2008}.

Deriving starting spectroscopic parameters from the main isotopologue permitted to fit H$_2$C$^{34}$S 
rotational data well with varying only a small subset of the parameters.

We have extended the set of rotational transition frequencies of thioketene to higher values of $J$ 
thus improving the spectroscopic parameters considerably with respect to the latest millimetre wave 
investigation \cite{H2CCS_rot_1980}.

\section*{Acknowledgement(s)}

We thank Don McNaughton for sending us the $\nu _2$ data of H$_2$CS and Jean-Marie Flaud for providing 
the 10$\umu$m line list prior to publication and the spectrum of H$_2$CS in a digital form more recently. 
We are grateful to Frank C. De Lucia for making equipment available to carry out the measurements at 
The Ohio State University and the late Manfred Winnewisser for assistance during these measurements. 
We thank Christian P. Endres and Monika Koerber for support during some of the measurements in K\"oln. 
We thank the Regionales Rechenzentrum der Universit{\"a}t zu K{\"o}ln (RRZK) for 
providing computing time on the DFG funded High Performance Computing System CHEOPS. 
Our research benefited from NASA's Astrophysics Data System (ADS).

\section*{Disclosure statement}

No potential conflict of interest was reported by the author(s).

\section*{Data availability statement}

The line, parameter and fit files with information on the setup of the parameter file, the transition 
frequencies with uncertainties, quantum numbers and residuals between observed frequencies and 
those calculated from the spectroscopic parameters, the parameters with codes, values and uncertainties 
and finally the correlation coefficients are available as supplementary material to this article 
together with an explanatory file. These files are all regular text files.  

These and auxiliary files are also available in the Cologne Database for Molecular Spectroscopy 
(CDMS) \cite{CDMS_2005,CDMS_2016} at https://cdms.astro.uni-koeln.de/classic/predictions/daten/H2CS/2023/. 
Calculations of the rotational spectra are available in the catalogue section of the CDMS at 
https://cdms.astro.uni-koeln.de/classic/entries/.

\section*{Funding}

We acknowledge support by the Deutsche Forschungsgemeinschaft via the collaborative 
research centers SFB~494 project E2 and SFB~956 (project ID 184018867) project B3 
as well as the Ger{\"a}tezentrum SCHL~341/15-1 (``Cologne Center for Terahertz Spectroscopy''). 
We are grateful to NASA for its support of the OSU program in laboratory astrophysics 
and the ARO for its support of the study of large molecules.




\bibliographystyle{tfo}
\bibliography{H2CS}


\end{document}